\documentclass[12pt]{article}
\usepackage{a4}
\def\taa{{t_a}}
\def\tbb{{t_b}}
\def\ta2{{t_a^2}}
\def\tb2{{t_b^2}}
\usepackage{amsmath}
\usepackage{epsfig}
\usepackage[german, english]{babel}
\def\beq{\begin{equation}}
\def\eeq{\end{equation}}
\def\beqn{\begin{eqnarray}}
\def\eeqn{\end{eqnarray}}
\begin{document}
\selectlanguage{english}
\title {The ($\gamma^* \to q\bar{q}$) - Reggeon  Vertex
  in Next-to-Leading Order QCD
  \unitlength1pt
  \begin{picture}(0,0)
    \put(100,80){
      {\hfill\small\parbox{3cm}{\raggedleft hep-ph/0009102\\DESY 00-132}}}
  \end{picture}}
%\\[0.5cm]
%\end{Large}
\author{J. Bartels, S. Gieseke, C.-F. Qiao \\[9pt]
  II. Institut f\"ur Theoretische Physik, Universit\"at
  Hamburg\\
  Luruper Chaussee 149, 22761 Hamburg, Germany}
%\end{center}
%
\date{  }
\maketitle
\vskip 9mm
\begin{abstract}
  As a first step towards the computation of the NLO corrections to
  the photon impact factor in the $\gamma^*\gamma^* \to \gamma^*
  \gamma^*$ scattering process, we calculate the one loop corrections 
  to the coupling of
  the reggeized gluon to the $\gamma^*\to q\bar{q}$ vertex. We list
  the results for the Feynman diagrams which contribute: all loop
  integrations are carried out, and the results are presented in the
  helicity basis of photon, quark, and antiquark.
\end{abstract}
%\vspace{2mm}

\hspace{4mm}PACS number(s): 11.55.Jy, 12.38.Bx, 13.60.-r

\vskip 15mm
\section{Introduction}

The experimental test of the BFKL Pomeron \cite{bfkl} is generally
considered to be an important task in strong interaction physics.
Recently much interest has been given to the total cross section of
the scattering of two highly virtual photons  $\sigma_{tot}^{\gamma^*
\gamma^*}$ \cite{bartels2,brodsky}. 
%,brodsky,bartels2,wallon,bialas,nikolaev,levin,dosch,kwiecinskiz,DSR,
This process describes the scattering of two small-size projectiles,
and its high energy behavior (at not too large energies) is expected
to be described by the BFKL Pomeron.  Therefore, a measurement of the
reaction $e^+e^- \to e^+e^- + X$ by tagging the outgoing leptons at
LEP or at a future Linear Collider provides an excellent test of this
very important QCD prediction.

So far leading order calculations of the BFKL Pomeron have been
compared to LEP data (both OPAL and L3) \cite{bartels1, DSR, NLOint}.
In both experiments the data lie above the one gluon exchange curve
(commonly called Born approximation), but below the BFKL prediction.
Since the next-to-leading order (NLO) corrections to the BFKL kernel
have been calculated \cite{FL,CC}, it is known that the higher
corrections will lower the theoretical predictions of the cross
section.  However, a consistent comparison with the NLO BFKL
calculations has not yet been possible: there remains the task of
calculating also the next-to-leading order corrections of the coupling
of the BFKL Pomeron to the external photons, the so-called photon
impact factor.

\begin{figure}
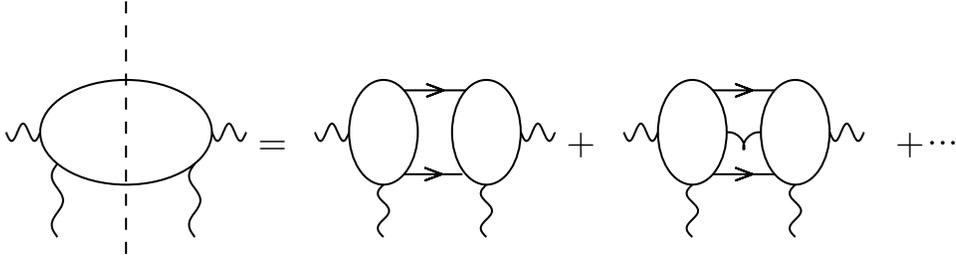

\begin{center}
\input fig1.pstex_t
\end{center}
\caption{Contributions to the photon impact factor.}
\label{fig:intermedstates}
\end{figure}
The photon impact factor is obtained from the energy discontinuity of
the amplitude $\gamma^* + \mbox{\textsl{reggeon}} \to \gamma^* +
\mbox{\textsl{reggeon}}$ (Fig.~1).  In leading order $\alpha_s$ this 
discontinuity
is simply the square of the scattering amplitude $\gamma^* +
\mbox{\textsl{reggeon}} \to q\bar{q}$ in the tree approximation, and
the reggeon, i.e. the reggeized gluon, can be identified with the
elementary t-channel gluon (with a particular helicity). In the
next-to-leading order new contributions have to be calculated.  For
the $q\bar{q}$ intermediate state we need the NLO corrections to the
$\gamma^* + \mbox{\textsl{reggeon}} \to q\bar{q}$ amplitude either on
the lhs or on the rhs of the discontinuity line, and the $q \bar{q} g$
intermediate state requires with leading order amplitudes $\gamma^* +
\mbox{\textsl{reggeon}} \to q\bar{q}g$ on both sides of the energy 
discontinuity line. The task of calculating the NLO corrections to the 
photon impact
factor therefore can therefore be organized in three steps, (i) the
calculation of the NLO corrections to the $\gamma^* +
\mbox{\textsl{reggeon}} \to q\bar{q}$ vertex, (ii) the vertex
$\gamma^* \to q\bar{q}g$ in leading order, and (iii) the integration
over the phase space of the intermediate states. In this paper we
report on results of the first step, the NLO corrections to the
$\gamma^* + \mbox{\textsl{reggeon}} \to q\bar{q}$ vertex. The vertex is 
obtained from the high energy limit of the scattering process 
$\gamma^* + q \to q\bar{q} + q$. 

\section{Technical preliminaries} 

\begin{figure}
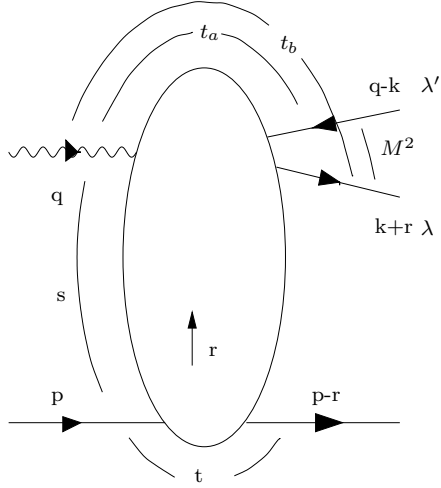

\begin{center}
\input fig2.pstex_t
\end{center}
\caption{Kinematics of the process $\gamma^* + q \to q\bar{q} + q$.}
\label{kinematics}
\end{figure}

The kinematics is illustrated in Fig.~2. As usual, $q$ and $p$ denote 
the four momenta of the photon and the incoming quark, resp., and
$\varepsilon_{L,t}$ the polarization vectors of the photon. 
We use $s$ to denote the energy
of the $\gamma^* q$ scattering process process, and we introduce the invariants
$Q^2 = -q^2$, $t_a = k^2$, $t_b = (q - k - r)^2$, $M^2 = (q + r)^2$, 
$t = r^2$, and $x = Q^2/2p\cdot{q}$ for the Bjorken scaling variable.
For simplicity, in the calculation of this paper we treat the quarks as 
massless. The momenta $k$ and $r$ can be written in the Sudakov 
decomposition form, i.e.,
\begin{figure}[t]
  \begin{center}
    \epsfig{file=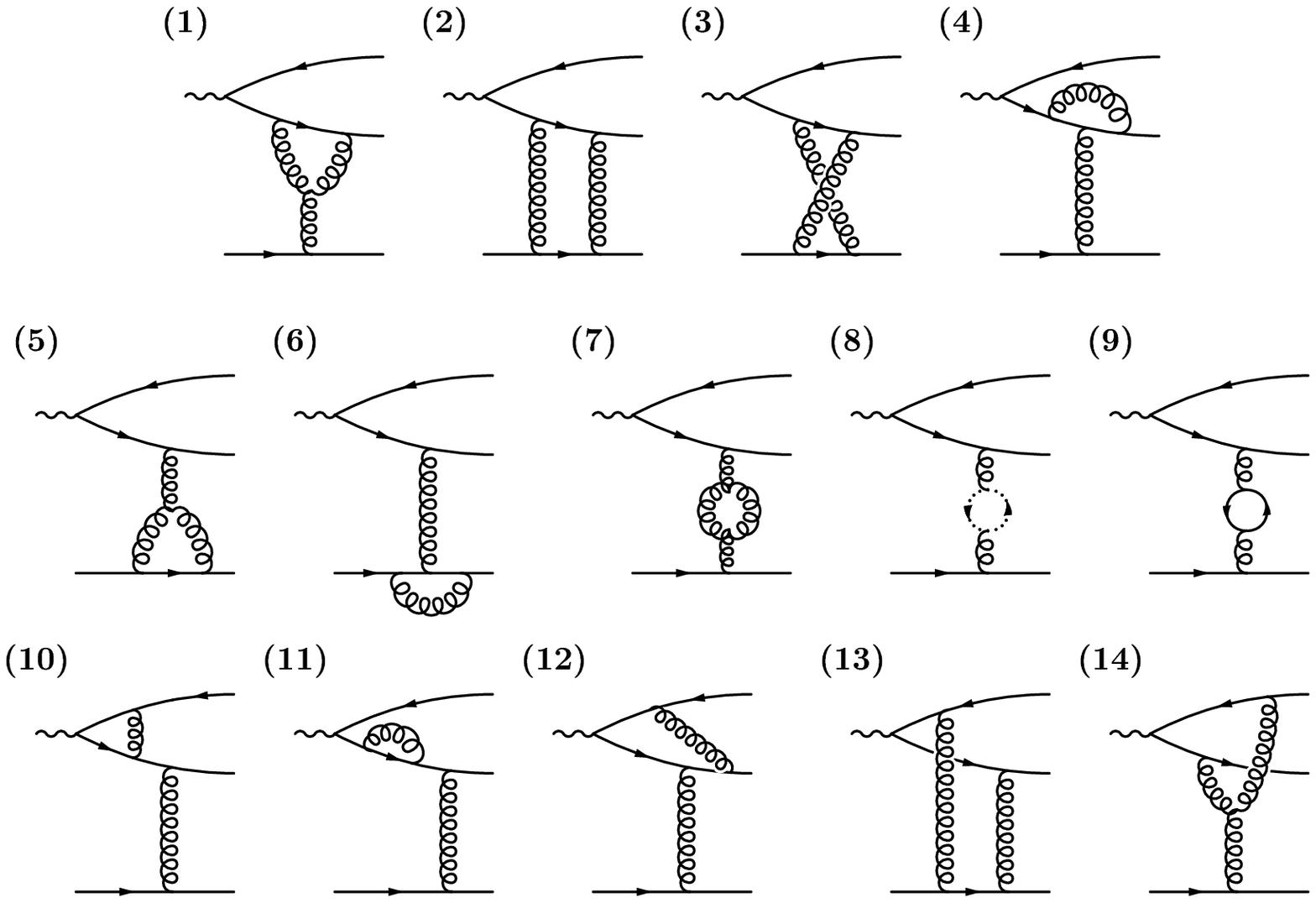,width=6in}
  \end{center}
  \caption{Feynman diagrams for the process $\gamma^* + q \to q\bar{q} +
    q$.}
  \label{graphs}
\end{figure}
\beqn
k = \alpha q' + \beta p + k_{\perp}\; ,
\eeqn
\beqn
r =  \frac{t}{s} q' - \frac{t_a + t_b}{s} p + r_{\perp}\; ,
\eeqn 
where $q' = q + x p$ and 
\beqn 
\beta s = \frac{k_{\perp}^2}{1 - \alpha} - Q^2. 
\eeqn
The Feynman diagrams which contribute to our NLO-calculation are listed in 
Fig.~3. In addition to the graphs shown (all diagrams, except for Fig.~3.14), 
we have to add those diagrams where the t-channel gluon couples to the 
outgoing antiquark rather than the outgoing quark.
It is easy to see that, for the color octet $t$-channel configuration, the sum 
of all diagrams has to be antisymmetric if we interchange quark and
antiquark: $k \to q - k - r$, $\lambda \to \lambda'$. In particular, 
the 'box' graph shown in Fig.~3.14 has to be antisymmetric by itself.
We will use the Feynman gauge throughout the calculation, and for the
t-channel gluons we decompose the metric tensor according to  
\beqn
\label{gmunu}
g_{\mu\nu}=\frac{2}{s} (p_{\mu} q_{\nu}'+ p_{\nu} q_{\mu}') + 
g_{\mu\nu}^{\perp}.
\eeqn
In our calculation we retain only the first term, since the 
remaining ones are suppressed by powers of the energy.
We use the helicity formalism, and our results will be expressed in terms of 
the following matrix elements:
\beqn
\label{HTdef}
H_{T}^a&=& \bar{u}(k+r, \lambda) \not\!{p} \not\!{k} \not\!{\varepsilon}\; 
\lambda^av(q-k,\lambda') \; , \\
\label{HTbardef}
\bar{H}_T^a&=&
\bar{u}(k+r, \lambda) \not\!{\varepsilon}\;  (\not\!{q}-\not\!{k}-\not\!{r})\; 
\not\!{p}\; \lambda^av(q-k,\lambda') \; , \\
\label{HEdef}
H_{\varepsilon}^a&=& \bar{u}(k+r, \lambda) \not\!{\varepsilon}\;  \lambda^a 
v(q-k,\lambda') \; , \\
\label{HKdef}
H_{k}^a&=& \bar{u}(k+r, \lambda) \not\!{k}\; \lambda^a v(q-k,\lambda')
\; , \\
\label{HPdef}
H_{p}^a&=& \bar{u}(k+r, \lambda) \not\!{p}\; \lambda^a v(q-k,\lambda')
\; ,
\eeqn 
where $\lambda^a(a = 1, \cdots, 8)$ are the color matrices, and $\lambda$ 
and $\lambda'$ denote the helicities of the outgoing quark pair. 
When interchanging quark and antiquark lines it will be convenient to use 
the following identity
\beqn
H_T^a + \bar{H}_T^a = s H_{\varepsilon}^a - 2 \varepsilon \cdot p H_{k}^a - 
2 \varepsilon \cdot r H_{p}^a \; .
\eeqn  
We organize our calculations in the following order. We first consider 
those diagrams (Fig.~3.1--9) which can be viewed as ``elastic scattering of 
two quarks'', with the upper `incoming' quark carrying the mass $t_a$.
Correspondingly, the diagrams not shown in Fig.~3 define quark-quark 
scattering with the upper `incoming' antiquark having the mass $t_b$. 
Together with the correction at the photon vertex (Fig.~3.10) and the quark 
self-energy (Fig.~3.11), these diagrams are the ones which have to be
made ultraviolet finite by renormalization. In the final part we turn
to the calculation of the box diagrams in Figs.~3.12 and 3.14, and 
the pentagon graph shown in Fig.~3.13. 

We are interested in the high energy limit, where
\beqn
\label{HElimit}
t,\; Q^2,\; t_a,\; t_b,\; M^2 \; \ll s \; ,
\eeqn 
and we do not impose any restriction on the remaining invariants.        
The Regge ansatz for the scattering amplitude 
$\gamma^*q \to (q\bar{q})q$ takes the form:
\beqn
\label{ansatz1}
T = \Gamma_{\gamma^* \to q\bar{q}}^a \frac{s}{t} 
\left[\left( \frac{\quad s}{-t} \right)^{\omega} + \left( \frac{-s}{-t}
  \right)^{\omega} \right] \Gamma_{qq}^a \; .
\eeqn
Here, $1 + \omega$ is the gluon trajectory. Expanding all terms in 
powers on the strong coupling $g$, we have
\beqn
\omega &=& g^2\;\omega^{(1)} + g^4\;\omega^{(2)} \; ,\\[3pt]
\Gamma_{\gamma^* \to q\bar{q}}^a&=&
g\;\Gamma_{\gamma^* \to q\bar{q}}^{(0),a} + 
g^3\;\Gamma_{\gamma^* \to q\bar{q}}^{(1),a} \; ,\\[3pt]
\Gamma_{qq}^a &=& g\; \Gamma_{qq}^{(0),a} + g^3\;\Gamma_{qq}^{(1),a}\; .
\eeqn
After substituting corresponding elements in Eq.(\ref{ansatz1}) with above
expansions, one obtains the following structure of the amplitude
\beqn
\label{eq:16}
T =  g^2\;T^{(0)} + g^4\;T^{(1)} 
\eeqn
with
\beqn
T^{(0)} = \Gamma_{\gamma^* \to q\bar{q}}^{(0),a}\;
\frac{2s}{t} \; \Gamma_{qq}^{(0),a} 
\eeqn
and 
\beqn
\label{expansion}
T^{(1)} &=&\Gamma_{\gamma^* \to q\bar{q}}^{(1),a}\; \frac{2s}{t}\;
\Gamma_{qq}^{(0),a} + \Gamma_{\gamma^* \to q\bar{q}}^{(0),a}\; 
\frac{2s}{t} \Gamma_{qq}^{(1),a}\nonumber\\ 
&&+\,\Gamma_{\gamma^* \to q\bar{q}}^{(0),a}\;
\frac{s}{t} \, \omega^{(1)} 
\left[ \ln \frac{{\quad s}}{-t} + \ln \frac{-s}{-t} \right] 
\Gamma_{qq}^{(0),a}.
\eeqn
Here, the $e_f$ represents the charge value of the interacting quark,
and the lowest-order expressions on the rhs of (\ref{expansion}) are
\beqn
\label{born1}
\Gamma_{qq}^{(0),a}&=& \frac{1}{s} \bar{u}(p-r, \,\lambda_{q'}) \not\!{q}' 
\lambda^a u(p,\,\lambda_q)\,, \\[3pt]
\label{lotraj}
\omega^{(1)}(t) &=& \frac{2 N_c}{(4\pi)^{2-\epsilon}}
\,\frac{c_{\Gamma}}{\epsilon}\,(-t)^{-\epsilon}\,, \\[3pt]
\label{born3}
\Gamma_{\gamma^* \to q\bar{q}}^{(0),a} &=& -ie_fe \left(\frac{H_T^a}{st_a} - 
\frac{\bar{H}_T^a}{st_b} \right)\,,
\eeqn
with $D=4 - 2\;\epsilon$ and
\beqn
c_{\Gamma} 
= \frac{\Gamma(1 + \epsilon)\Gamma^2 (1 - \epsilon)}
{\Gamma(1-2\epsilon)}\approx 1-\gamma_E\epsilon
+\frac{1}{2}\left(\gamma_E^2-\frac{\pi^2}{6}\right)\epsilon^2 + {\cal O}
(\epsilon^3)\,   
\eeqn 
Here $\lambda_q$ and $\lambda_{q'}$ are the helicities of the incoming
and outgoing quark, resp., and $\lambda^a$ are the generators of the
colour group. In this paper we will present the results of Figs.~3.1 --
3.14 which can be cast into the form of the rhs of (\ref{expansion}).
All pieces except for $\Gamma_{\gamma^* \to q\bar{q}}^{(1),a}$ are
known from earlier calculations.  In particular, the Born result for
$\Gamma_{\gamma^* \to q\bar{q}}^{(0),a}$ is well known in the context
of the photon wave function formalism, and it is available for
definite helicity states \cite{wavefun}.  The higher order corrections
to the quark-quark-reggeon vertex, $g^3 \Gamma_{qq}^{(1)}(t)$, have
been calculated in \cite{fadin2}. For vanishing quark masses these
corrections are (following the notations of \cite{fadin2}):
\begin{equation}
  \Gamma_{qq}^{(1), a}(t) = \lambda^a\left(\Gamma_{qq}^{(1)}
    (q\bar q-\mbox{state}) +
    \Gamma_{qq}^{(1)}(gg-\mbox{state})\right)
  \delta_{\lambda_{q}\lambda_{q'}},  
\end{equation}
with
\begin{align}
  \label{NLOqqvertqq}
  \Gamma_{qq}^{(1)}(q\bar q-\mbox{state}) & =
  \frac{(-t)^{-\epsilon}}{(4\pi)^{2-\epsilon}}\left[
  -\frac{n_f}{2}\,\frac{c_\Gamma}{\epsilon(1-2\epsilon)}
  \left(1-\frac{1}{3-2\epsilon}\right) 
  +\frac{c_\Gamma}{2N_c}\left(\frac{2}{\epsilon^2}
    +\frac{3}{\epsilon}+8\right)
  \right]\\
  \label{NLOqqvertgg}
  \Gamma_{qq}^{(1)}(gg-\mbox{state}) & = 
  \frac{N_c\,(-t)^{-\epsilon}}{(4\pi)^{2-\epsilon}}\left[
  -\frac{c_\Gamma}{\epsilon^2} + \frac{1}{3}\frac{c_\Gamma}{\epsilon}
  +\frac{13}{18}+\frac{\pi^2}{2}
  \right]. 
\end{align}
Inserting these results on the rhs of (\ref{expansion}) we can
easily obtain $\Gamma_{\gamma^* \to q\bar{q}}^{(1),a}$, which is 
the goal of this paper. 

For future purposes it will be important to note that the vertex
$\Gamma_{\gamma^* \to q\bar{q}}^{(1),a}$ in (\ref{ansatz1}) is
expected \cite{bartels3} to have a rather complex structure. For
example, in the limit of large diffractive masses it will be more
convenient to write (\ref{ansatz1}) as a sum of two different
expressions: the first one depending on $M^2$ and $s$, the second one
on $\alpha s$ and $s$ (or $(1-\alpha)s$).  This decomposition exhibits
the reggeization of both the gluon and the quark. A detailed
discussion of the large-$M^2$ limit will be presented in a seperate
paper \cite{BGQ2}. 

\section{Analytic Results}
We write the NLO amplitude $T^{(1)}$ for the process $\gamma^*+q\to q\bar q+q$
as a sum of the different Feynman diagrams: 
\beqn
\label{sumAi}
T^{(1)} = \sum_{i = 1}^{13} \big(A_i + \bar{A_i}) + A_{14}. 
\eeqn
The subscripts $i$ refer to the
numbering in Fig.~3, and the amplitudes $\bar{A_i}$ correspond to the
diagrams which are not shown in Fig.~3: they are obtained by interchanging 
the couplings of $t$-channel gluons between quark and antiquark lines.
Formally, we substitute
\begin{eqnarray}
&&k\leftrightarrow q-k-r, \;\alpha \leftrightarrow ( 1 - \alpha ),\;
t_a \leftrightarrow t_b, \nonumber\\
&&H^a_T \leftrightarrow \bar{H}^a_T, \;
\varepsilon\cdot k \leftrightarrow -\varepsilon\cdot (k + r),\;
\lambda\leftrightarrow\lambda', 
\end{eqnarray}
Under these replacements the matrix elements
(\ref{HEdef})--(\ref{HPdef}) remain unchanged. In addition, for all
diagrams except for Fig.~3.14 we have to include an overall minus sign
when interchanging quark and antiquark. $A_{14}$ is already
antisymmetric by itself with respect to the interchange of quark and
antiquark.

\subsection{Calculational methods}
Before presenting explicit analytic expressions for all diagrams
we briefly outline the methods we have used to obtain the results. 

Starting from the standard Feynman rules of QCD we project the color
in the $t$-channel into the antisymmetric octet and proceed by
introducing Feynman parameters $x_i$ to combine the denominators for
the integration of the loop momentum. Only for the simplest diagrams
these steps are easily performed by hand. Particularly for the
diagrams Fig.~3.12--14 this step becomes too tedious. Therefore, we have
used the computer algebra system Mathematica with the package FeynCalc
\cite{feyncalc}, in order to reduce the numerators to expressions
which contain the helicity matrix elements (\ref{HTdef}) to
(\ref{HPdef}), monomials $x_i x_j\cdots$ in the Feynman parameters
and, of course, the kinematical invariants.  Those diagrams, where the
loop integral itself does not depend on the large scale $s$, the box
diagrams Fig.~3.12 and Fig.~3.14, have to be calculated exactly. The
only high energy approximation results from the gluon nonsense
helicity (i.e.\ the first term in (\ref{gmunu}). Diagram Fig.~3.13, on the
other hand, has first been calculated exactly, and then the high
energy limit (\ref{HElimit}) has been taken.

To consider the loop integrations, we had to deal with integrals over
the loop momentum itself, which could be easily performed by the usual shift, 
and with the remaining integrals of the Feynman parameters. Integrals of the 
latter type are either known, or they could be obtained from known ones 
with the help of recurrence relations to ${\cal O}(\epsilon^0)$ in dimensional regularization. The
technical background for these methods is given in \cite{bern} and has
been applied to almost all cases we are interested in by
\cite{durham}. We have written a Mathematica package to easily call
the results of \cite{durham}. In most cases the integrals are
recursively expressed in terms of special functions, representing a
particular combination of logarithms and dilogarithms for a given
$n$-point function. In the case of diagram Fig.~3.14 we have calculated
explicit results for integrals with two and three Feynman parameters
in the numerator using the derivative method \cite{bern}.  Also for
this task we have used Mathematica. 

Finally, after carrying out the integrals, i.e.\ replacing the
monomials of Feynman parameters in our amplitudes with the explicit
expressions from our Mathematica package, we have used again FeynCalc to
take the high energy limit in the diagrams Figs.~3.2, 3, and 13, and to
carry out some simplifying algebra. A few final simplifications 
had to be done by hand. The expressions that we have obtained
in this way will be listed in the remainder of this section.

\subsection{Quark-quark Scattering}

It is suggestive to view the diagrams Figs.~3.1--9
as a quark-quark scattering processes with one of the incoming quarks being 
off-shell (with virtuality $t_a$). We focus on those diagrams in 
Fig.~\ref{graphs}, where the $t$-channel gluon(s) couples to the quark. 
Those diagrams where the gluon(s) couples to the antiquark are easily obtained 
by performing the substitutions described after (\ref{sumAi}).

Diagram Fig.~3.1 is one of many three point functions with two massive
external legs. Loop integrals of this type are well-known, and we have
calculated them both by hand and by using computer algebra as outlined
above.
\begin{align}
\label{a1}
A_1\,&=\, \frac{N_c}{2} \frac{1}{(4 \pi)^{2 - \epsilon}}
\bigg\{ A^{(0)} 
  \bigg[
    \frac{2c_\Gamma}{\epsilon^2} \frac{t_a}{t-t_a} 
    \left((-t_a)^{-\epsilon}- (- t)^{-\epsilon}\right) \\
    &+ \frac{c_\Gamma}{\epsilon} (- t)^{-\epsilon} + 2   +
    \frac{2 t}{t - t_a}{\rm{ln}}\frac{t}{t_a} \bigg] - 
  \frac{(-iee_f) \alpha}{t-t_a} H^a_{\varepsilon}\frac{2 s}{t} 
  \Gamma_{qq}^{(0),a} \nonumber\\
&\times \bigg[
    - \frac{c_\Gamma}{\epsilon^2} \frac{t_a}{t-t_a} 
    \left((-t_a)^{-\epsilon}- (- t)^{-\epsilon}\right) 
+ \frac{ c_\Gamma}{\epsilon} (-t)^{-\epsilon} + 1  
- \frac{t}{t - t_a}{\rm{ln}}\frac{t}{t_a}\bigg] \bigg\} \; .
\nonumber
\end{align}
Here and in the following, we use 
\begin{equation}
A^{(0)} = - i e e_f \frac{H_T^a}{s t_a}\,\frac{2 s}{t}\,
\Gamma_{qq}^{(0),a}
\end{equation}
as a shorthand notation. 

In the diagrams Fig.~3.2 and Fig.~3.3 we have to deal with four-point
integrals, where one of the external legs is off-mass shell. Here we
make use of the shorthand notations listed in the appendix. The result
for the sum of the two diagrams is:
\begin{align}
A_{2+3} =& - A^{(0)} N_c \frac{1}{(4 \pi)^{2 - \epsilon}}
\left \{\frac{c_\Gamma}{\epsilon^2} \Big[
(\alpha s)^{- \epsilon} + (- \alpha s)^{- \epsilon} + 
(- t)^{- \epsilon} - 2 (- t_a)^{-\epsilon}\right.\nonumber \\
&-\left.
\frac{t}{t - t_a}\left( (- t)^{-\epsilon} - ( - t_a)^{- \epsilon} \right)
\Big] + {\rm{Ld}_0^{1m}}(\alpha s, t, t_a) + 
{{\rm{Ld}}_0^{1m}} (-\alpha s, t, t_a) \right\}
\end{align}
with
\begin{eqnarray}
{\rm{Ld}}_0^{1m}(\alpha s, t, t_a)  = {\rm{Li}}_2 (1 - \frac{t}{t_a} ) +
{\rm{Li}}_2 (1 - \frac{\alpha s }{t_a} ) +
{\rm{ln}}\frac{t}{t_a}{\rm{ln}}\frac{\alpha s}{t_a}- \frac{\pi^2}{6}\; .
\nonumber
\end{eqnarray}
Here, $\rm{Li}_2$ is the standard dilogarithm function defined as
\begin{eqnarray}
{\rm{Li}}_2(x) = - \int^1_0 \frac{\ln (1 - x t)}{t} dt \;. 
\end{eqnarray}
In order to exhibit the energy dependence we rewrite this expression:
\begin{align}
\label{logsform}
A_{2+3} =& 
- A^{(0)} \frac{N_c}{(4 \pi)^{2 - \epsilon}} 
\Big\{ 
-\frac{c_\Gamma(-t)^{-\epsilon}}{\epsilon} 
\left[ \ln \frac{{\alpha s}}{-t} + \ln \frac{-\alpha s}{-t} \right] 
\nonumber\\
&+ 
2\,{\rm Li}_2\left(1-\frac{t}{t_a}\right) 
+ \left(\ln\frac{t}{t_a}\right)^2 +\frac{\pi^2}{3}\nonumber+
(-t)^{-\epsilon}\left[\frac{2c_\Gamma}{\epsilon^2} -\pi^2\right]\nonumber\\
&+\frac{c_\Gamma}{\epsilon^2} \Big[
(- t)^{- \epsilon} - 2 (- t_a)^{-\epsilon}
- \frac{t}{t - t_a}\left( (- t)^{-\epsilon} - ( - t_a)^{- \epsilon} \right)
\Big]
\Big\}. 
\end{align}
From these two diagrams (and from their respective partners $\bar
A_{2+3}$) we already get the complete $\ln s$ dependence of our
result, in agreement with the rhs of (\ref{expansion}). In detail, the
first line of (\ref{logsform}) contains, apart from the $\ln \alpha $
dependence, the leading order trajectory function $\omega^{(1)}(t)$
(cf.(\ref{lotraj})). Furthermore, if we take the limit $t_a\to 0$ in
the last two lines, we are left with only the last term of the second
line.  One half of this term will go to the lower vertex
(\ref{NLOqqvertgg}) (the same result could also have been deduced from
~\cite{fadin2}).  Therefore, the contribution from diagrams 2 and 3 to
$\Gamma_{\gamma^* \to q\bar{q}}^{(1),a}$ is just
\begin{align}
\label{contrib2and3}
& i e e_f \frac{H_T^a}{s t_a}\frac{N_c}{(4 \pi)^{2 - \epsilon}} 
\Big\{ 
-\frac{2 c_\Gamma(-t)^{-\epsilon}}{\epsilon} 
\ln \alpha 
+ 
2\,{\rm Li}_2\left(1-\frac{t}{t_a}\right) 
+ \left(\ln\frac{t}{t_a}\right)^2 +\frac{\pi^2}{3} \nonumber\\[3pt]
&+
\frac{(-t)^{-\epsilon}}{2}\left[\frac{2c_\Gamma}{\epsilon^2} 
  -\pi^2\right]
+\frac{c_\Gamma}{\epsilon^2} \Big[
(- t)^{- \epsilon} - 2 (- t_a)^{-\epsilon}
- \frac{t}{t - t_a}\left( (- t)^{-\epsilon} - ( - t_a)^{- \epsilon} \right)
\Big]
\Big\}. 
\end{align}
We emphasize, once more, that the two diagrams $A_{2+3} + \bar
A_{2+3}$ provide the complete $\ln s$ dependence on the rhs of
(\ref{ansatz1}). At first sight one might expect that also the
pentagon diagrams $A_{13}+\bar A_{13}$ might contribute to the energy
dependence. Later we will show that, in fact, this is not the case.

The calculation of Fig.~3.4 is very similar to the calculation of
Fig.~3.1. We find:
\begin{align}
\label{a4}
&A_4= - \frac{1}{2 N_c} \frac{1}{(4 \pi)^{2 - \epsilon}} 
\bigg\{ A^{(0)} 
 \bigg[\frac{2 c_\Gamma}{\epsilon^2} 
   \frac{t}{t-t_a} \left ( (-
     t_a)^{-\epsilon} - ( - t)^{- \epsilon} \right )
   - \frac{c_\Gamma}{\epsilon} (- t)^{-\epsilon} \\
   &\; - 4 + \frac{2 t + t_a}{t - t_a} {\rm{ln}}\frac{t}{t_a} 
 \bigg] -iee_f\, \alpha H^a_{\varepsilon} \frac{2s}{t} \Gamma_{qq}^{(0), a} 
\Big[ 
  \frac{2 c_\Gamma}{\epsilon^2} \frac{t \left(
    (-t_a)^{- \epsilon} - ( - t)^{- \epsilon} 
  \right)}{(t-t_a)^2} \nonumber\\
 &\; - \frac{2 c_\Gamma}{\epsilon} \frac{( - t)^{-
        \epsilon}}{t - t_a} + \frac{ 3 t + 2 t_a}{(t - t_a)^2}
    {\rm{ln}}\frac{t}{t_a} - \frac{5}{t - t_a} 
    \Big] 
\bigg\}\;.\nonumber
\end{align}

Diagrams Fig.~3.5 and Fig.~3.6 are needed to obtain
the NLO corrections to the lower vertex $\Gamma^{(1), a}_{qq}$.
In agreement with ~\cite{fadin2} we find: 
\begin{equation}
\label{a5}
A_5 = - A^{(0)}\;\frac{N_c c_\Gamma} 
{(4 \pi)^{2 - \epsilon}}\left(\frac{(-t)^{-\epsilon}}{2\epsilon} + 
1 \right)\; ,
\end{equation}
and
\begin{equation}
\label{a6}
A_6 = A^{(0)} \;\frac{1}{(4 \pi)^{2 - \epsilon}}\frac{
c_\Gamma}{2 N_c} (-t)^{-\epsilon} \left(\frac{2}{\epsilon^2} + 
\frac{3}{\epsilon} + {8}\right)\; .
\end{equation}
We note that they can be obtained from Eqs.~(\ref{a1}) and (\ref{a4}) 
by choosing $\epsilon < 0$ and taking the limit $t_a\to 0$.

The calculation of diagrams Figs.~3.7--9 is straightforward. They contribute
to both the upper and the lower vertex:
\begin{eqnarray}
\label{a7+8}
A_{7+8} = A^{(0)} \; \frac{N_c c_\Gamma}
{(4 \pi)^{2 - \epsilon}} \frac{(-t)^{-\epsilon}}{2 \epsilon 
(1 - 2 \epsilon)} \left(\frac{1}{3 - 2 \epsilon} + 3\right)\; ,
\end{eqnarray}
Similarly, the $q\bar q$-contribution to the gluon-self-energy in Fig.~3.9
leads to 
\begin{eqnarray}
\label{a9}
A_9 = A^{(0)} n_f \; \frac{c_\Gamma}
{(4 \pi)^{2 - \epsilon}} \frac{(-t)^{-\epsilon}}{\epsilon 
(1 - 2 \epsilon)} \left(\frac{1}{3 - 2 \epsilon} - 1\right). 
\end{eqnarray}
These diagrams contribute with equal weight to both the upper and to
the lower vertex. Therefore, in order to complete the upper
reggeon-quark-quark vertex with one off-shell quark, we simply add
$1/2$ of the sum of (\ref{a7+8}) and (\ref{a9}) to the contributions
(\ref{a1}), (\ref{a4}), and (\ref{a5}).  Similarly, the lower
reggeon-quark-quark vertex with all quarks being on-shell is obtained
by adding $1/2$ of the sum of (\ref{a7+8}) and (\ref{a9}) to
(\ref{a5}), (\ref{a6}) and the contribution of the box diagrams
Figs.~3.2 and 3.3:

\begin{equation}
-A^{(0)} \frac{N_c}{(4\pi)^{2-\epsilon}}  
\frac{(-t)^{-\epsilon}}{2}\left[\frac{2c_\Gamma}{\epsilon^2} 
  -\pi^2\right]. 
\end{equation}

To summarize, so far we have analyzed the diagrams Figs.~3.1--9. From
Figs.~3.5, 3.6, $1/2$ of Figs.~3.7--9, and from a piece of Fig.~3.2
and 3.3.  we have reproduced the lower reggeon-quark-quark vertex of
~\cite{fadin2}.  From Figs.~3.1, 3.4, $1/2$ of Figs.~3.7--9, and from
the major part of Figs.~3.2 and 3.3 we have computed a new
reggeon-quark-quark vertex with one quark having the mass $t_a$.
Moreover, from Figs.~3.2 and 3.3. we have extracted the $\ln s$ terms
of the rhs of (\ref{expansion}). The remaining diagrams Figs.~3.10--14
provide contributions to the upper vertex $\Gamma^{(1),a}_{\gamma^*\to
  qq}$ only.

\subsection{Vertex correction and quark self-energy}

In this subsection we present the results of the vertex correction 
Fig.~3.10:
\begin{align}
  \label{a10}
  A_{10} &= -  \frac{C_{\rm{F}}}{(4 \pi)^{2 - \epsilon}}
  \bigg\{A^{(0)} 
    \bigg[
    \frac{2 c_\Gamma}{\epsilon^2} 
    \frac{Q^2}{Q^2+t_a} \left ( (Q^2)^{-\epsilon} 
      - ( -t_a)^{- \epsilon} \right )
    + \frac{c_\Gamma}{\epsilon} (- t_a)^{-\epsilon} \\
  &+ 4 + \frac{3 Q^2}{Q^2 + t_a}{\rm{ln}}\frac{-t_a}{Q^2} 
    \bigg]
  -\Gamma' \frac{2 s}{t} \Gamma_{qq}^{(0), a} 
    \bigg[
    \frac{4 c_\Gamma}{\epsilon^2} 
    \frac{Q^2}{Q^2+t_a} \left ( (Q^2)^{-\epsilon} 
      - ( -t_a)^{- \epsilon} \right )\nonumber\\
    &+ \frac{4 c_\Gamma}{\epsilon} (- t_a)^{-\epsilon} 
    + 10\left(1 + \frac{Q^2}{Q^2 + t_a} \right) {\rm{ln}}\frac{-t_a}{Q^2}
    \bigg]
  \bigg\}\nonumber
\end{align}
with
\begin{equation}
\Gamma^{\prime} = - i e e_f\, \frac{H_p^a}{s} 
\frac{\varepsilon\cdot k}{Q^2 + t_a},
\end{equation}
and for the quark self-energy Fig.~3.11:
\beqn
\label{a11}
A_{11} = - A^{(0)} \frac{C_{\rm{F}}}{(4 \pi)^{2 -
\epsilon}} (-t_a)^{-\epsilon}\frac{c_\Gamma (1 - \epsilon)}{\epsilon 
(1 - 2 \epsilon)}\; . 
\eeqn
Eqs. (\ref{a10}) and (\ref{a11}) provide new contributions to the vertex 
$\Gamma^{(1),a}_{\gamma^*\to qq}$.

\pagebreak 
\subsection{The Box diagram Fig.~3.12}
In this subsection we give the result of the diagram Fig.~3.12, which
was calculated entirely with the help of the computer algebra (in
\cite{durham} this diagram has been named `adjacent box'). The result
will be expressed as follows:
\begin{equation}
  A_{12} = 
  \Gamma^{(0), a}_{qq} \frac{2\,s}{t} \left(\frac{-ie e_f}{s}
    \right) \left(-\frac{1}{2N_c}\right) 
  \frac{c_\Gamma}{(4\pi)^{2-\epsilon}}
  \left[\frac{1}{\epsilon^2} A_{12}^{(-2)} 
    + \frac{1}{\epsilon} A_{12}^{(-1)} + A_{12}^{(0)}\right]. 
\end{equation}
Starting with the divergent terms, we have
\begin{align}
A_{12}^{(-2)} &=  
  \frac{2\alpha\,H^a_\varepsilon\,s\,t\,
    ( {( -t ) }^{-\epsilon} - {( -t_a ) }^{-\epsilon} ) }{(t - t_a)^2} 
+ 2 H^a_T \bigg[\frac{{( -t ) }^{-\epsilon} - {( -t_a ) }^{-\epsilon}}{t
- t_a} \;\\ 
&+ \; \frac{{( -t_a ) }^{-\epsilon} - {(Q^2)}^{-\epsilon}}{Q^2 +
t_a} +\frac{-{( -M^2 ) }^{-\epsilon} + {(Q^2)}^{-\epsilon} 
+ {( -t ) }^{-\epsilon} - 2{( -t_a ) }^{-\epsilon}}{t_a} \bigg]\nonumber\\
&- \frac{4 Q^2 \varepsilon\cdot k H_p^a}{(Q^2+t_a)^2} 
  \left ( (Q^2)^{-\epsilon} - ( -t_a)^{- \epsilon} \right )
\nonumber
\end{align}
and 
\begin{align}
A_{12}^{(-1)} =&  
  4\,\epsilon\cdot k\,H^a_p \left[
  \frac{ 
    ( {( -t ) }^{-\epsilon} - 
    {( -t_a ) }^{-\epsilon} ) }{M^2} 
  - \frac{{( -t_a ) }^{-\epsilon}}{Q^2+t_a}
  \right] \nonumber\\
& +
  H^a_\varepsilon s\,\left[ 
  \frac{2\,\alpha (-t)^{-\epsilon}}{t - t_a}
  + \frac{( 1 - \alpha ) \,
    ( {( -t ) }^{-\epsilon}-{( -t_a ) }^{-\epsilon}) }{M^2}\right].
\end{align}
The ${\cal O}(\epsilon^0)$ term for the this adjacent box is expressed in
terms of many different functions, related to this loop integral. For
convenience, the definitions of these functions are listed in the
appendix. 
\begin{align}
A_{12}^{(0)} &=  
 -2 (2 H^a_T + s \alpha H^a_\varepsilon) {\rm Lc}_{0}(-Q^2, M^2, t)
\nonumber\\
& + 
\big\{3 H^a_T M^2 + H^a_k [6 \varepsilon\cdot p M^2 
 + 2 s ((8 \alpha - 3) 
\varepsilon\cdot k 
+ (5 \alpha - 3) \varepsilon\cdot r)] 
\nonumber \\
&\quad - 
s \alpha\,H^a_\varepsilon (3 t_a - 2 t_b - Q^2)  - 
2 H^a_p (2 Q^2 \varepsilon\cdot r + \varepsilon\cdot k (M^2 - Q^2 - 3
t_b)) \big\}\frac{{\rm Lc}_{1}(M^2, -Q^2, t)}{M^2}\nonumber \\ 
& +  
2 \big\{H^a_T M^2 + 2 H^a_k (M^2 \varepsilon\cdot p - s [\varepsilon\cdot k + 
(1 - \alpha)\varepsilon\cdot r]) + 2 s H^a_\varepsilon 
[(1 - \alpha) t + \alpha Q^2  + t_a]  \nonumber \\
&\quad -  
4 H^a_p [\varepsilon\cdot k (2 M^2 + Q^2 + t ) + \varepsilon\cdot
  r ( Q^2 + t_a )] \big\} \frac{{\rm Lc}_{1}(-Q^2,M^2,t)}{M^2}
  \nonumber\\
& + 
\big\{2 \varepsilon\cdot k H^a_p (2 M^2 - 2 Q^2 - t_a )  
 - ( 1 - \alpha ) s Q^2 H^a_\varepsilon - H^a_T ( Q^2 + t_a) \nonumber \\
&\quad- 
2 H^a_k [ \varepsilon\cdot k (1 - 2 \alpha ) s + 
\varepsilon\cdot p ( Q^2 + t_a )] \big\}\frac{{\rm
  Lc}_{1}^{2m}(-Q^2, t_a)}{M^2} \nonumber \\
& - 2 \big\{2 s \alpha (3 \varepsilon\cdot k + \varepsilon\cdot r)
  H^a_k + s H^a_\varepsilon [\alpha ( M^2 - Q^2 - 3 t_a )  + ( t - t_a )] 
\nonumber\\
&\quad + H^a_p [ 3 \varepsilon\cdot r t_a  
 +  \varepsilon\cdot k\,(4\,t - t_a
  )]\big\} \frac{{\rm Lc}_{1}^{2m}(t_a, t)}{M^2} \nonumber\\
& +  2 \big\{4 \varepsilon\cdot p M^2 \,H^a_k + 3 M^2 H^a_T 
 - s [\alpha ( M^2 - 2\,Q^2 - t )  + t ] H^a_\varepsilon \nonumber 
\end{align}
\begin{align}
&\quad - 2 [\varepsilon\cdot r\,Q^2 + \varepsilon\cdot k\,( M^2 + Q^2 + 2\,t )] 
H^a_p \big\}\frac{{\rm Lc}_{1S}(-Q^2,M^2,t)}{M^2 t_a} \nonumber\\
& + 4 \alpha s H^a_\varepsilon {\rm Lc}_{2}^{2m}(t_a,t) 
- 8\,\varepsilon\cdot k\,H^a_p {\rm Lc}_{2}^{2m}(-Q^2,t_a) 
\nonumber\\
& -2\,\big\{ H^a_T\,M^2 + 2 H^a_k[\varepsilon\cdot p M^2 
- s((1-2\alpha)\varepsilon\cdot k + (1-\alpha)\varepsilon\cdot r)] \nonumber\\
&\quad +2\, H_p^a\varepsilon\cdot k (2M^2 + t - t_a) - \alpha s
H^a_\varepsilon (Q^2+t_a)
\big\}\frac{{\rm Lc}_{2}(-Q^2,M^2,t)}{M^2} \nonumber\\
&+  2 \big\{2 \varepsilon\cdot k (t_a - M^2) H^a_p + ( Q^2 + t_a) H^a_T
\nonumber\\
&\quad + 2 [( 1 - 2\,\alpha ) s \varepsilon\cdot k + \varepsilon\cdot p ( Q^2
+ t_a )] H^a_k \big\} \frac{{\rm Lc}_{3}(M^2,-Q^2,t)}{M^2} \nonumber\\
& - 
2 (2\varepsilon\cdot k\,H^a_p + s \alpha\,H^a_\varepsilon )
{\rm Lc}_{3}(-Q^2,M^2,t) - 4\,H^a_T \frac{{\rm Ld}_{0}^{a}
(t_a, M^2, -Q^2, t)}{t_a} \nonumber\\
& + 
2 \big\{ t_b M^2 H^a_T + s t_a [\alpha\,( Q^2 + t ) - 
Q^2 - 2\,\alpha\,M^2 - t_b] H^a_\varepsilon \nonumber\\
&\quad - 2 t_a [\varepsilon\cdot k\, ( 2 M^2 + Q^2 + t) 
+ \varepsilon\cdot r\,(t - t_b)] H^a_p  -
2 [ ( -( \varepsilon\cdot p\,M^2 ) + ( \varepsilon\cdot k + 
\varepsilon\cdot r ) s) t_a \nonumber \\
&\quad + s \alpha (2\,\varepsilon\cdot k\,M^2 - 
\varepsilon\cdot r\,t_a )] H^a_k \big\} 
\frac{{\rm Ld}_{1}(t_a,M^2,-Q^2,t)}{M^2 t_a} \nonumber \\
& - 
  2 \{H^a_T  + s\,H^a_\varepsilon - 2 \,\varepsilon\cdot r\,H^a_p \}
{\rm Ld}_{1S}(t_a, M^2, -Q^2, t) 
\nonumber\\
&+ 
\big\{4 [\alpha\,s\,( 6\,\varepsilon\cdot k\, (t - t_b) + 
             2\,\varepsilon\cdot k\,t_a + 
             2\,\varepsilon\cdot r\,Q^2 
 + 3\,\varepsilon\cdot r\,t_a ) \nonumber \\ 
&\quad - t_a ( \varepsilon\cdot p\,M^2 + 
              s\,( \varepsilon\cdot k + \varepsilon\cdot r ))] H^a_k  
 + 4 [ 2\,\varepsilon\cdot r\,t_a\, (t - t_b)  + 
\varepsilon\cdot k\,( 2 t (Q^2 + t) \nonumber \\
&\quad + 
(t - t_a ) (t_a - 2 t_b) )] H^a_p 
- 2\,s [ \alpha (Q^2 +  t_a) (2\,Q^2 + 5 t_a)  - 2 ( Q^2 t + t_a t_b) 
]\,H_\varepsilon^a \nonumber \\
&\quad - 2 H^a_T M^2 t_a \big\} \frac{{\rm Ld}_{21}(t_a, M^2, -Q^2, t)}
{2 t_a M^2} \nonumber \\
& +
 2 \big\{ - ( Q^2 + t_a ) M^2 H^a_T + 
    2 [ \varepsilon\cdot r\, ( Q^2 + t_a ) \, ( t - t_b ) \nonumber \\
&\quad + \varepsilon\cdot k\,( (Q^2 + t_a) (t - t_a) + (t - t_b)^2) ]
H^a_p - 
2 [  \varepsilon\cdot p\,M^2\, ( Q^2 + t_a )   \nonumber\\
&\quad - 
s\,( ( 1 - \alpha ) \,\varepsilon\cdot r\,( Q^2 + t_a )   
- \varepsilon\cdot k\,(Q^2  + t_a - (1 - 3\,\alpha)\,M^2 ) ) ] H^a_k
\nonumber \\
&\quad + 
s\,[ ( 1 - \alpha) (Q^2 t + t_a t_b) 
- \alpha\,( Q^2 + t_a )^2 ] H^a_\varepsilon \big\}
\frac{ {\rm Ld}_{22}(t_a,M^2,-Q^2,t)} {M^2\,t_a} \nonumber \\
& +
  \frac{2\,s\,\alpha\,( Q^2 + 2\,t_a ) \, H^a_\varepsilon\,
     {\rm Ld}_{24}(t_a,M^2,-Q^2,t)}{t_a} 
\nonumber\\
& + \big\{ 24\,\varepsilon\cdot p M^2 \,H^a_k + 18\,M^2 H^a_T
 - 6\,s [ t - t_a  - \alpha\, (3 \,Q^2 + 2 t_a + t_b) ] H^a_\varepsilon 
\nonumber \\
&\quad - 12 [\varepsilon\cdot k\,( 3\,t - 2 t_a - t_b)  + 
\varepsilon\cdot r\,( Q^2 + t_a )  \
) ] H^a_p \big\}\frac{{\rm Ld}_{2S} (t_a, M^2, -Q^2, t)}
{M^2 t_a} \nonumber \\
& +
\big\{8 \varepsilon\cdot k\,t_a (t_a - t) H^a_p - 4 ( Q^2 t + t_a t_b )\,
H_T^a 
+ 8 [ \varepsilon\cdot k\,s\,(t_a - t )  - \alpha\,\varepsilon\cdot k\,s\,
( M^2 - 2\,t + 2\,t_a )   \nonumber \\
&\quad -  \varepsilon\cdot p\,( Q^2\,t + t_a\,t_b ) ]
H^a_k \big\}  \frac{{\rm Ld}_{311}(t_a, M^2, -Q^2, t)} {2 M^2 t_a}\nonumber \\
& + \big\{ 4\,\varepsilon\cdot k\, {( t - t_a ) }^2 H^a_p 
 - 4\,s\, [ (\varepsilon\cdot k + \varepsilon\cdot r ) \,
( t - t_a )  - \alpha\,( 2\,\varepsilon\cdot k + \varepsilon\cdot r \
) \,( Q^2 + t_b ) ] H^a_k \nonumber \\
&\quad - 2\,\alpha\,s\, ( Q^2\,t + t_a t_b) \,H^a_\varepsilon\big\}
   \frac{{\rm Ld}_{314}(t_a,M^2,-Q^2,t)}{M^2 t _a}  \nonumber \\
& +
4\,( 1 - \alpha )\, s \,\varepsilon\cdot k\,
H^a_k\, \frac{{\rm Ld}_{322}(t_a, M^2,-Q^2,t)}{t_a} - 4\,\alpha \,s\,( \varepsilon\cdot k + \varepsilon\cdot r ) \,
H^a_k\,\frac{{\rm Ld}_{344}(t_a,M^2, -Q^2,t)}{t_a} . 
\end{align}
%\oddsidemargin=-0.2in
%\end{document}

%%% Local Variables: 
%%% mode: latex
%%% TeX-master: "paper71"
%%% End: 

\pagebreak
\subsection{The Box diagram Fig.~3.14}
The result of the diagram Fig.~3.14 (in ~\cite{durham} named `opposite
box'), can, again, be split into divergent and finite pieces:
\begin{equation}
  A_{14} = 
  \Gamma^{(0), a}_{qq} \frac{2s}{t} \left(\frac{-ie e_f}{s}\right) 
  \left(\frac{N_c}{2}\right) 
  \frac{1}{(4\pi)^{2-\epsilon}}
  \left[\frac{c_\Gamma}{\epsilon^2} A_{14}^{(-2)} 
    + \frac{c_\Gamma}{\epsilon} A_{14}^{(-1)} + A_{14}^{(0)}\right]. 
\end{equation}
The term proportional to $c_\Gamma/\epsilon^2$ reads: 
\begin{align}
A_{14}^{(-2)} &= \bigg\{
\frac{2\,\,\varepsilon\cdot p\,H^a_k\,\left( t_a - t_b \right) }{Q^
2\,t + t_a\,t_b} + \frac{\,(H^a_T - \bar H^a_T)\,
\left( t_a + t_b \right) }{Q^2\,t + t_a\,t_b} \nonumber\\ 
&\quad +
\frac{\,H^a_\varepsilon\,s\,}{2 
        {\left( t - t_a \right) }^2\,
        {\left( t - t_b \right) }^2\,
        \left( Q^2\,t + t_a\,t_b \
\right) }
\Big[ (\alpha\, - ( 1 - \alpha)) \left( Q^2\,t + t_a\,t_b \right) \nonumber\\
&\quad\quad\times 
  \left( t^2\,\left( t_a + t_b \right)  + 
   t_a\,t_b\, \left( t_a + t_b \right) - 4\,t\,t_a\,t_b  \right)\nonumber\\  
&\quad\quad -
\left( t_a - t_b \right)\,
\left( 2 (t - t_a)^2 ( t - t_b)^2  + 
\left( Q^2\,t + t_a\,t_b \right) \left(t_a t_b - t^2\right) \right)
\Big] \nonumber\\
&\quad - 
\frac{2\,\,H^a_p\,}{\left( t - t_a \right) \,
        \left( t - t_b \right) \,
        \left( Q^2\,t + t_a\,t_b \right) }
\Big[ \varepsilon\cdot ( k + r) \,
 \left( t - t_b \right) \, \left( t_a\,\left( t_a - t_b \right)  + 
             t\,\left( t_a + t_b \right) \right)\nonumber\\  
&\quad\quad + \varepsilon\cdot k\,\left( t\, - t_a \right)
   \left(t_b \left( t_b - t_a \right) \,
              + t\,\left( t_a + t_b \right) \right) \Big] \bigg\} 
{\left( -t \right) }^{-\epsilon} \nonumber  \\
%\end{align}
%\vspace{-5mm}
%%%%%%%%%%%%%%%%%%%%%%%%%%%%
%\begin{align}
& + 
\bigg\{\frac{-2\,\,
        \varepsilon\cdot p\,H^a_k\,M^2\,Q^2\,
        \left( t_a - t_b \right) }{\left({Q^2} + t_a \right) \,
        \left( Q^2 + t_b \right) \,
        \left( Q^2\,t + t_a\,t_b \
\right) } + \frac{\,H^a_\varepsilon\,M^2\,
        Q^2\,s\,\left( t_a - t_b \
\right) }{\left( Q^2 + t_a \right) \,
        \left( Q^2 + t_b \right) \,
        \left( Q^2\,t + t_a\,t_b \
\right) } \nonumber\\ 
&\quad + \frac{\,(H^a_T - \bar H^a_T)\,Q^2\,
        \left( {t_a}^2 + {t_b}^2 - 
          t\,\left( t_a + t_b \right)  + 
          Q^2\,\left(t_a + 
             t_b - 2\, t \right)  \right) }{\left( Q^2 + 
          t_a \right) \,
        \left( Q^2 + t_b \right) \,
        \left( Q^2\,t + t_a\,t_b \
\right) } \nonumber\\ 
&\quad -
\frac{2\,\,H^a_p\,Q^2\,
         }{{\left( Q^2 + 
            t_a \right) }^2\,
        {\left( Q^2 + t_b \right) }^2\,
        \left( Q^2\,t + t_a\,t_b \right) }
\Big[ \varepsilon\cdot k\,\left( Q^2 + t_b \right) \,
 \Big( Q^4\, \left( t_a - t_b \right)  \nonumber\\
&\quad\quad +  Q^2\,
\left( 2\,{t_a}^2 + t_a\,t_b - {t_b}^2 + 
t\,\left( t_b - 3 t_a \right) \right) 
+ t_a\,\left( {t_a}^2 + {t_b}^2 - t\,\left( t_a + t_b \right)  \
\right)  \Big)  \nonumber\\  
&\quad\quad + 
\varepsilon\cdot (k + r ) \,
\left( Q^2 + t_a \right) \,
\Big( Q^4\, \left( t_b - t_a \right)  
+ Q^2\,\left( 2 t_b^2  + t_a t_b - {t_a}^2 + t (t_a - 3 t_b )\right) 
\nonumber\\
&\quad\quad + 
t_b\,\left( {t_a}^2 + {t_b}^2 - t\,\left( t_a + t_b \right)  \right) 
 \Big)  \Big] \bigg\}
{\left(Q^2\right)^{-\epsilon}} \nonumber\\
%\end{align}
%%%%%%%%%%%%%%%%%%%%%%%%%%%%
%\begin{align}
& +
\bigg\{\frac{\,(H^a_T - \bar H^a_T)\,
        \left( Q^4 + M^2\,Q^2 - {t_a}^2 \right) }{\left( Q^2 + 
 t_a \right) \, \left( Q^2\,t + t_a\,t_b \
\right) } - \frac{2\,\,\varepsilon\cdot p\,H^a_k\,
 [ {t_a}^2 + Q^2\,\left( t + t_a - t_b \
\right) ] }{\left( Q^2 + t_a \right) \, \left( Q^2\,t + t_a\,t_b \
\right) } \nonumber\\
&\quad - 
\frac{\,H^a_\varepsilon\,s\, }{{\left( t - t_a \right) }^2\,
        \left( Q^2 + t_a \right) \,
        \left( Q^2\,t + t_a\,t_b \
\right) }\Big[ \alpha \,t_a (Q^4 t + t_a^2 t_b)\, -
t_a^2 (t - t_a)^2   \nonumber\\
&\quad\quad +
Q^2\,\left( -t^3 + t^2 \,(t_a\, + t_b) + t\,t_a (t_a - 
2\,t_b )  + t_a^2\,\left( t_b - t_a \right)  + \alpha {t_a}^2\,
\left( t + t_b \right)  \right)  \Big]  \nonumber
\end{align}
\begin{align}
&\quad +
\frac{2\,\,H^a_p\,}
{\left( t - t_a \right) \,{\left( Q^2 + t_a \right) }^2\,
\left( Q^2\,t + t_a\,t_b \right) }
\Big[ \varepsilon\cdot k\,\left( t - t_a \right) \,
\big( {Q^6} + M^2\,Q^2\,\left( Q^2 - t_a \right)  \nonumber\\
&\quad\quad + {Q^4}\,t_a + Q^2\,{t_a}^2 + {t_a}^3 \big) 
+ \varepsilon\cdot ( k + r ) \, \left( Q^2 + t_a \right) \,
\big( {t_a}^2\, \left( t + t_a \right)  \nonumber\\ 
&\quad\quad + Q^2\,\left(t_a\, \left( t_a - t_b \right) + 
t\,\left( 2\,t_a + t_b \right)  - t^2 \right)  \big)  \Big]\bigg\}
{\left( -t_a \right) }^{-\epsilon} \nonumber\\
%\end{align}
%%%%%%%%%%%%%%%%%%%%%%%%%%%%
%\begin{align}
& +
\bigg\{\frac{\,(H^a_T - \bar H^a_T)\,
        \left( Q^4 + M^2\,Q^2 - {t_b}^2 \right) }{\left( Q^2 + 
 t_b \right) \, \left( Q^2\,t + t_a\,t_b \
\right) } + \frac{2\,\,\varepsilon\cdot p\,H^a_k\,
 [ {t_b}^2 + Q^2\,\left( t + t_b - t_a \
\right) ] }{\left( Q^2 + t_b \right) \, \left( Q^2\,t + t_a\,t_b \
\right) } \nonumber\\
&\quad - 
\frac{\,H^a_\varepsilon\,s\, }{{\left( t - t_b \right) }^2\,
        \left( Q^2 + t_b \right) \,
        \left( Q^2\,t + t_a\,t_b \
\right) }\Big[ -(1 - \alpha) \,t_b (Q^4 t + t_b^2 t_a)\, +
t_b^2 (t - t_b)^2   \nonumber\\
&\quad\quad +
Q^2\,\left(t^3 - t^2 \,(t_a\, + t_b) + t\,t_b (2 t_a - 
t_b )  + t_b^2\,\left( t_b - t_a \right)  - (1 - \alpha) {t_b}^2\,
\left( t + t_a \right)  \right)  \Big]  \nonumber\\
&\quad +
\frac{2\,\,H^a_p\,}
{\left( t - t_b \right) \,{\left( Q^2 + t_b \right) }^2\,
\left( Q^2\,t + t_a\,t_b \right) }
\Big[ \varepsilon\cdot (k + r)\,\left( t - t_b \right) \,
\big( {Q^6} + M^2\,Q^2\,\left( Q^2 - t_b \right)  \nonumber\\
&\quad\quad + {Q^4}\,t_b + Q^2\,{t_b}^2 + {t_b}^3 \big) 
+ \varepsilon\cdot k  \, \left( Q^2 + t_b \right) \,
\big( {t_b}^2\, \left( t + t_b \right)  \nonumber\\
&\quad\quad + Q^2\,\left(t_b\, \left( t_b - t_a \right) + 
t\,\left( 2\,t_b + t_a \right)  - t^2 \right)  \big)  \Big]\bigg\}
{\left( -t_b \right) }^{-\epsilon} .
\end{align}
%%%%%%%%%%%%%%%%%%%%%%%%%%%%%%%%%%%%%%%%%%%%%%%%%%%%%%%%%%%%%%%%%%%%%%%%%%%
The $c_\Gamma/\epsilon$ - term is rather simple, since we kept only
the first terms in the expansion of powers like $(-t)^{-\epsilon}$ while
the logarithms are combined with logarithms from the finite term,
leading to significant simplifications:  
\begin{align}
A_{14}^{(-1)} = 
\frac{s \,H^a_\varepsilon [ \alpha ( t - t_b) - (1 - \alpha) ( t - t_a )
]} {\left( t - t_a \right) \, \left( t - t_b \right) } -
\frac{4\,H^a_p [\varepsilon\cdot k\,\left( Q^2 + t_b \ \right)
+ \varepsilon\cdot (k + r)\,\left(\,Q^2 + t_a \right) ]}{\left( Q^2 +
  t_a \right) \, \left( Q^2 + t_b \right) }. 
\end{align}
%%%%%%%%%%%%%%%%%%%%%%%%%%%%%%%%%%%%%%%%%%%%%%%%%%%%%%%%%%%%%%%%%%%%%%%%%%%
Finally, the ${\cal O}(\epsilon^0)$ term for this opposite box diagram reads:
\begin{align}
A_{14}^{(0)} &= 
\frac{2\,s\, [\alpha (t - t_b) 
  - ( 1 - \alpha) (t -t_a)  ] \,H^a_\varepsilon}{\left( t - t_a \right) \,
  \left( t - t_b \right) } - 
\frac{8\, [\varepsilon\cdot (r + k) \left( Q^2 + t_a \right)  + 
  \varepsilon\cdot k\,\left( Q^2 + t_b \right) ] H^a_p}
{\left( Q^2 + t_a \right) \, \left( Q^2 + t_b \right) } \nonumber \\
& +
\frac{2\,s\, [\alpha\,\varepsilon\cdot ( k + r )\, \left( Q^2 + t_a \right) \,
\left( t - t_b \right)  + ( 1 - \alpha ) \varepsilon\cdot k\,
( Q^2 + t_b ) \left( t - t_a \right) ] H^a_k} 
{ ( t - t_a ) \,( t - t_b )\,( Q^2 + t_a )  \, ( Q^2 + t_b ) } \nonumber\\
& -
\frac{{\rm Ld}_0^{op}(t_a, t_b, -Q^2, t)}{{M^4}\,\left( Q^2\,t + 
       t_a\,t_b \right) } 
   \bigg\{ (H^a_T - \bar H^a_T)\,M^2\,
        \Big[ 3 (Q^2 t + t_a\,t_b ) + M^2 ( t_a + t_b ) \Big] \nonumber\\ 
&\quad + 2\,H^a_k\,\Big[ \varepsilon\cdot p\,M^4\,
\left( t_a - t_b \right)  - 3\,s\,
\left( Q^2\,t + t_a\,t_b \right) (\alpha\,\varepsilon\cdot k +
\left( 1 - \alpha \right) \,\varepsilon\cdot (k + r) )\Big]  
\nonumber\\
&\quad + H^a_p\,\Big[ \varepsilon\cdot r\,(t_a - t_b) 
  \left( 2Q^4 +Q^2 (4
    (t_a + t_b) - 5t)- 2t(t_a+t_b) +2(t_a+t_b)^2 -3t_a\,t_b\right)\nonumber\\ 
&\quad\quad +
\left( \varepsilon\cdot k + \varepsilon\cdot (k + r) \right) \,
\big( 4 t {Q^4} + Q^2 t ( 9 (t_a + t_b) - 10 t ) 
-2 t^2 (t_a + t_b) \nonumber\\
&\quad\quad + 2 t ( t_a^2 + t_b^2 - t_a t_b) + 3 t_a t_b(t_a + t_b)
\big) \Big] \nonumber
\end{align}
\begin{align}
&\quad -
\frac{H^a_\varepsilon\,s}{2}\,\Big[ (t_a - t_b) ( 2 Q^4 - 7 Q^2 t + 4
Q^2 (t_a + t_b) + 2 ( t - t_a - t_b)^2 - 3 t_a t_b )  \nonumber\\ 
&\quad\quad - 3\, [\alpha - (1 - \alpha)] (Q^2\,t\, + t_a t_b)
\left(2 Q^2 + t_a + t_b \right)  \big]\bigg\} \nonumber\\
%%%%%%%%%%%%%%%%%%%%%%%%%%%%%%%%%%%%
& - 
\bigg\{ \frac{3\,H^a_\varepsilon\,Q^2\,s}
{M^2\,\left( Q^2 + t_a \right) \,\left( Q^2 + t_b \right) } 
\Big[ \alpha (Q^2 + t_a) - ( 1 - \alpha) ( Q^2 + t_b ) \Big] \nonumber\\  
&\quad -
\frac{2\,H^a_p\,Q^2}{M^2\,{\left( Q^2 + t_a \
\right) }^2\,{\left( Q^2 + t_b \right) }^2}
\Big[ \varepsilon\cdot k\,
\left( 4 M^2 + 3\,( Q^2 + t_a)\right) (Q^2 + t_b)^2 \nonumber\\
&\quad\quad +
\varepsilon\cdot (k + r) \left( 4 M^2 + 3 (Q^2 + t_b)\right) 
(Q^2 + t_a)^2 \Big] \nonumber\\
&\quad - 
\frac{H^a_k\,Q^2\,s}{M^2\,
        {\left( Q^2 + t_a \right) }^2\,
        {\left( Q^2 + t_b \right) }^2\,
        \left( Q^2\,t + t_a\,t_b \right) }
\Big[ \varepsilon\cdot k ( Q^2 + t_b)^2 \big( 6 \alpha ( Q^2 + t_a) (
        t - t_b) \nonumber\\
&\quad\quad + 
2 ( 1 - \alpha) t_a M^2 \big) 
+ \varepsilon\cdot (k + r) \, ( Q^2 + t_a )^2\,
\left( 6\,\left( 1 - \alpha \right) \,
\big( Q^2 + t_b \right) ( t - t_a) \nonumber\\
&\quad\quad + 
2 \alpha t_b M^2 \big)
\Big] 
\bigg\} \,\log (Q^2) \nonumber\\
%%%%%%%%%%%%%%%%%%%%%%%%%%%%%%%%%%%
& + 
\bigg\{ \frac{(H^a_T - \bar H^a_T)\,
          (2\,t - t_a - t_b )}{2\,\left( t - t_a \right) \,
        \left( t - t_b \right) } 
+ 
\frac{H^a_p\,}{M^2\,\left( t - t_a \right) \,
\left( t - t_b \right) }\Big[ \varepsilon\cdot (k + r) 
\big( M^2 ( t_b - t_a) \nonumber\\
&\quad\quad - 
6 t ( t - t_b) \big)  -
\varepsilon\cdot k\,\left(M^2 ( t_b - t_a) + 6 t ( t - t_a) \right)
\Big] \nonumber \\
&\quad +
\frac{H^a_\varepsilon\,s\,}{2\,M^2\,{\left( t - t_a \right) }^2\,
{\left( t - t_b \right) }^2}\Big[ t\,\left( t_a - t_b \right) \,
\big( (4 t -2\,t_a  - 2 t_b) Q^2 - t^2 \nonumber\\  
&\quad\quad + 
3 t (t_a + t_b) - 2 t_a^2 - 2 t_b^2 - t_a t_b \big) \nonumber\\
&\quad\quad + (1 - \alpha) \big(-4 t^3 ( t_a + t_b) + t_a t_b (t_a + t_b)^2
  + t^2 (7 t_a^2 + 6 t_a t_b + 7 t_b^2) \nonumber\\  
&\quad\quad - t (3 t_a^3 + 5 t_a^2 t_b + 5 t_a t_b^2 + 3 t_b^3)
\nonumber\\
&\quad\quad + 
Q^2 ( -6 t^3 + 7 t^2 ( t_a + t_b) + t_a t_b (t_a + t_b) -
 t ( 3 t_a^2 + 4 \taa \tbb + 3 \tb2) ) \big) \nonumber\\
&\quad\quad - 
\alpha \,\big(-4 t^3 ( t_a + t_b) + t_a t_b (t_a + t_b)^2
  + t^2 (7 t_a^2 + 6 t_a t_b + 7 t_b^2) \nonumber\\  
&\quad\quad - t (3 t_a^3 + 5 t_a^2 t_b + 5 t_a t_b^2 + 3 t_b^3) \nonumber\\
&\quad\quad + 
Q^2 ( -6 t^3 + 7 t^2 ( t_a + t_b) + t_a t_b (t_a + t_b) -
t ( 3 t_a^2 + 4 \taa \tbb + 3 \tb2) ) \big)\Big] \nonumber\\
&\quad -
\frac{H^a_k\,}{M^2\,{\left( t - t_a \right) }^2\,
{\left( t - t_b \right) }^2\,\left( Q^2\,t + t_a\,t_b \right) }
  \Big[\varepsilon\cdot p\,M^2\, \left( t - t_a \right) \,
  \left( t - t_b \right) \,\left( t_a - t_b \right) \,
   \left( Q^2\,t + t_a\,t_b \right) \nonumber\\
&\quad\quad +
\varepsilon\cdot k\,s\,t\,\left( t - t_a \right) \,
\big( 2\,M^2\,t_b\, (1 - \alpha) \left( t - t_a \right) +
6 \alpha (Q^2 + \tbb) (t - \tbb)^2 \big) \nonumber\\
&\quad\quad +
\varepsilon\cdot (k + r)\,s\,t\,\left( t - t_b \right)
\big( 2\,M^2\,t_a\, \alpha \left( t - t_b \right) +
6 (1 - \alpha) (Q^2 + \taa) (t - \taa)^2 \big) \Big\}
\log(-t)\nonumber\\
%%%%%%%%%%%%%%%%%%%%%%%%%%%
& + \bigg\{ \frac{H^a_T - \bar H^a_T}
{2(\taa - t)} + \frac{H^a_\varepsilon\,s\,}{2\,M^2\,
{\left( t - t_a \right) }^2\,\left( Q^2 + t_a \right) }
\Big[ M^2\,\left( Q^2 + t_a \right) (\taa - t + 4 \alpha t)
\nonumber\\  
&\quad\quad -
6 \taa ( t - ta) ( \taa + \alpha Q^2 - (1 - \alpha) t) \Big] +
\frac{H^a_p\,}{M^2\,\left( t - t_a \right)\,{\left( Q^2 + t_a \right)}^2}
\Big[ \varepsilon\cdot r\,M^2\,{\left( Q^2 + t_a \right) }^2\,
\nonumber\\
&\quad\quad + 6\,t_a \,\varepsilon\cdot r\,\left( Q^2 + \taa \right)^2  
+ 6\,t_a \,\varepsilon\cdot k\,\left( Q^2 + \taa \right) 
\left( Q^2 + t \right) + 4 \varepsilon\cdot k\,M^2 (Q^2 - \taa) (\taa
-t) \Big] \nonumber
\end{align}
\begin{align}
&\quad +  
\frac{H^a_k\,}{M^2\,{\left( t - t_a \right) }^2\,
        {\left( Q^2 + t_a \right) }^2\,
        \left( Q^2\,t + t_a\,t_b \right) } 
\Big[ \varepsilon\cdot p\,M^2\, \left( t - t_a \right) \,
{\left( Q^2 + t_a \right) }^2\,
\left( Q^2\,t + t_a\,t_b \right)  \nonumber\\
&\quad\quad +
s\,\varepsilon\cdot r\,\left( 2\,\alpha M^2 t + 6\,
(1 - \alpha) {\left( t - t_a \right) }^2 \right)\,t_a\,{\left( Q^2 + t_a \
\right) }^2 +
s\,\varepsilon\cdot k\,\big( 2\,M^2 t_a [\alpha t 
(Q^2 + t_a)^2  \nonumber\\
&\quad\quad - 
Q^2 (t - t_a)^2 - 2 \alpha Q^2 (t - t_a) (t +  2 t_a)]
+ 6 (t - t_a) [t_a (t - t_a) (Q^2 + t_a)^2 \nonumber\\
&\quad\quad - \alpha M^2 (Q^4 t + t_a^3)]
\big)\Big] \bigg\} \log(-t_a) \nonumber\\
%%%%%%%%%%%%%%%%%%%%%%%%
& + 
\bigg\{ \frac{H^a_T - \bar H^a_T}
{2(\tbb - t)} + \frac{H^a_\varepsilon\,s\,}{2\,M^2\,
{\left( t - t_b \right) }^2\,\left( Q^2 + t_b \right) }
\Big[ M^2\,\left( Q^2 + t_b \right) (t - \tbb - 4 (1 - \alpha) t) \nonumber\\
&\quad\quad -
6 \tbb ( t - t_b) (\alpha t - \tbb -(1 - \alpha) Q^2) \Big] + 
\frac{H^a_p\,}{M^2\,\left( t - t_b \right)\,{\left( Q^2 + t_b \right)}^2}
\Big[ -\varepsilon\cdot r\,M^2\,{\left( Q^2 + t_b \right) }^2\,\nonumber\\
&\quad\quad - 
6\,t_b \,\varepsilon\cdot r\,\left( Q^2 + t_b \right)^2 
+ 6\,t_b \,\varepsilon\cdot (k + r)\,\left( Q^2 + \tbb \right) 
\left( Q^2 + t \right) \nonumber\\ 
&\quad\quad + 
4 \varepsilon\cdot (k + r)\,M^2 (Q^2 - \tbb) (\tbb
-t) \Big] \nonumber\\
&\quad +  
\frac{H^a_k\,}{M^2\,{\left( t - t_b \right) }^2\,
        {\left( Q^2 + t_b \right) }^2\,
        \left( Q^2\,t + t_a\,t_b \right) } 
\Big[ -\varepsilon\cdot p\,M^2\, \left( t - t_b \right) \,
{\left( Q^2 + t_b \right) }^2\,
\left( Q^2\,t + t_a\,t_b \right)  \nonumber\\
&\quad\quad +
s\,\varepsilon\cdot r\,\left( -2\,(1 - \alpha) M^2 t - 6 \,
\alpha {\left( t - t_b \right) }^2 \right)\,t_b\,{\left( Q^2 + t_b \
\right) }^2 \nonumber\\ 
&\quad\quad + 
s\,\varepsilon\cdot (k + r)\,\big( 2\,M^2 t_b [(1-\alpha) t 
(Q^2 + t_b)^2  - Q^2 (t - t_b)^2 - 2 (1-\alpha) Q^2 (t - t_b) (t +  2
t_b)]\nonumber\\
&\quad\quad + 
6 (t - t_b) [t_b (t - t_b) (Q^2 + t_b)^2 - (1-\alpha) M^2 (Q^4 t + t_b^3)]
\big) \Big]\bigg\} \log(-t_b). 
\end{align}
We have cast all our results for $A_{14}$ into a form which manifestly
exhibits the antisymmetry under the exchange of $q$ and $\bar q$.

%%% Local Variables: 
%%% mode: latex
%%% TeX-master: "paper71"
%%% End: 

\subsection{The Pentagon diagram Fig.~3.13}
The result for the pentagon diagram Fig.~3.13, is much simpler than
one might expect. The reason for this is the fact that, in contrast to
the box diagrams Fig.~3.12 and 3.14, the integrals depend upon the
large scale $s$, and in the high energy limit they can enormously be
simplified.  As before, we seperate divergent and finite pieces:
\begin{equation}
  A_{13} = 
  \Gamma^{(0), a}_{qq} \frac{2s}{t} \left(\frac{-ie e_f\,t}{s}\right) 
  \left(\frac{N_c}{2}\right) 
  \frac{c_\Gamma}{(4\pi)^{2-\epsilon}}
  \left[\frac{1}{\epsilon^2} A_{13}^{(-2)}+
  A_{13}^{(0)}\right]. 
\end{equation}
The divergent contribution has the form: 
\begin{align}
\label{pentpole}
&A_{13}^{(-2)}  = 
\frac{1}{t} \left[\frac{H^a_T}{t_a} +\frac{\bar H^a_T}{t_b} \right]
\left({( \alpha\,s ) }^{-\epsilon} 
- {(- ( 1 - \alpha ) \,s ) }^{-\epsilon} \right) \nonumber\\
&\quad + 
\bigg\{
  \frac{t_a\,t_b}{t (Q^2\,t + t_a\,t_b)} 
    \left[\frac{H^a_T}{t_a} - \frac{\bar H^a_T}{t_b} \right]
  - \frac{2 H^a_T}{t\,t_a} 
  + \frac{2\,H^a_p}{Q^2\,t + t_a\,t_b}
  \left[\frac{\varepsilon\cdot k\,Q^2}{Q^2+t_a}
    -\frac{\varepsilon\cdot (k+r)\,t_a}{t-t_a}\right]
\bigg\}
{( -t_a ) }^{-\epsilon} \nonumber\\
&\quad + 
\bigg\{
  \frac{t_a\,t_b}{t (Q^2\,t + t_a\,t_b)} 
    \left[\frac{H^a_T}{t_a} - \frac{\bar H^a_T}{t_b} \right]
  + \frac{2 \bar H^a_T}{t\,t_b} 
  - \frac{2\,H^a_p}{Q^2\,t + t_a\,t_b}
  \left[\frac{\varepsilon\cdot k\,t_b}{t-t_b}
    -\frac{\varepsilon\cdot (k+r)\,Q^2}{Q^2+t_b}\right]
\bigg\}
{( -t_b ) }^{-\epsilon} \nonumber\\
&\quad + 
\bigg\{
  -\frac{t_a\,t_b}{t (Q^2\,t + t_a\,t_b)} 
    \left[\frac{H^a_T}{t_a} - \frac{\bar H^a_T}{t_b} \right]
  + \frac{2\,H^a_p}{Q^2\,t + t_a\,t_b}
  \left[\frac{\varepsilon\cdot k\,t_b}{t-t_b}
    +\frac{\varepsilon\cdot (k+r)\,t_a}{t-t_a}\right]
\bigg\}
{( -t ) }^{-\epsilon}\nonumber\\ 
&\quad + 
\bigg\{
  \frac{Q^2}{Q^2\,t + t_a\,t_b} 
    \left[\frac{H^a_T}{t_a} - \frac{\bar H^a_T}{t_b} \right]
  - \frac{2\,Q^2\,H^a_p}{Q^2\,t + t_a\,t_b}
  \left[\frac{\varepsilon\cdot k}{Q^2 + t_a}
    +\frac{\varepsilon\cdot (k+r)}{Q^2+t_b}\right]
\bigg\}
{Q^2}^{-\epsilon}, 
\end{align}
whereas the finite piece reads:
\begin{align}
\label{pentfin}
A^{(0)}_{13} & = 
\frac{t_a\,t_b}{t (Q^2\,t + t_a\,t_b)}
\left[\frac{H^a_T}{t_a}
  +\frac{\bar H^a_T}{t_b} \right]\,
  \Big[
  {\rm Ld^{a}_0}(t_a, (1 - \alpha ) \,s,-Q^2, - \alpha\,s  )\nonumber\\
&  - {\rm Ld^{a}_0}(t_b,- \alpha\,s, -Q^2,( 1 - \alpha ) \,s) 
 - 
{\rm Ld^{1m}_0}(( 1 - \alpha ) \,s, t, t_b) 
  + {\rm Ld^{1m}_0}(- \alpha\,s, t, t_a)
  \Big] \nonumber\\
& -  
\frac{2 H^a_T}{t\,t_a} {\rm Ld^{a}_0}(t_a, (1 - \alpha ) \,s,-Q^2, - \alpha\,s)
+\frac{2 \bar H^a_T}{t\,t_b}{\rm Ld^{a}_0}(t_b,-\alpha\,s, -Q^2,( 1 -
\alpha)\,s)  \nonumber\\
& + 
\bigg\{\frac{t_a\,t_b}{t (Q^2\,t + t_a\,t_b)} 
\left[\frac{H^a_T}{t_a} - \frac{\bar H^a_T}{t_b} \right] \nonumber\\
& - 
\frac{2\,H^a_p\,}{M^2\,\left( Q^2\,t + t_a\,t_b \right) } 
\Big[\varepsilon\cdot k\,(Q^2+t_b) + \varepsilon\cdot (k+r)\,(Q^2+t_a)\Big]
\bigg\} \,{\rm Ld^{op}_0}(t_a,t_b,t,-Q^2). 
\end{align}

%%% Local Variables: 
%%% mode: latex
%%% TeX-master: "paper71"
%%% End: 

The counterpart $\bar A_{13}$ of the pentagon graph Fig.~3.13 can be
obtained in two different ways. Either we follow the substitution
described after (\ref{sumAi}) and simply interchange quark and
antiquark $q\leftrightarrow \bar q$. Alternatively, we start from
Fig.~3.13 and perform the crossing $(s\to u\approx -s)$ (an additional
minus sign comes from the color antisymmetry in the $t$-channel).  In
the following we present the sum $A_{13}+\bar{A}_{13}$. In order to
demonstrate the cancellation of the $\ln s$-dependence we combine the
first line of (\ref{pentpole}) with the first three lines of
(\ref{pentfin}) (and their counterparts in $\bar{A}_{13}$).  The
results contain a term proportional to $c_{\Gamma}/\epsilon$:
\begin{align} 
   \frac{1}{t} \left[\frac{H^a_T}{t_a} +\frac{\bar H^a_T}{t_b} \right]
  \big[&\ln( (1-\alpha)\,s ) + \ln(- ( 1 - \alpha ) \,s ) \nonumber\\
    &-\ln( \alpha\,s ) - \ln(- \alpha\,s )\big].  
\end{align}
and a finite piece:
\begin{align}
  &\frac{t_b\,H^a_T+t_a\,\bar H^a_T}{t (Q^2\,t + t_a\,t_b)}
  \Big[
  \frac{1}{2}\big(\ln( (1-\alpha)\,s ) 
  - \ln(- ( 1 - \alpha ) \,s )\big)^2 
  -\frac{1}{2}\big(\ln( \alpha\,s ) 
  - \ln(- \alpha\,s )\big)^2 \nonumber\\
  &\quad+\big(\ln(-t_a) + \ln(-t_b)-\ln(-t) - \ln(Q^2)\big)\nonumber\\
  &\qquad \times\big(\ln( (1-\alpha)\,s ) + \ln(- ( 1 - \alpha ) \,s ) 
  -\ln( \alpha\,s ) - \ln(- \alpha\,s )\big)
  \Big]\nonumber\\
%%%%
  &+\frac{H^a_T}{t\,t_a}
  \Big[
  \frac{1}{2}\big(\ln( \alpha\,s ) 
  - \ln(- \alpha\,s )\big)^2
  - \frac{1}{2}\big(\ln( (1-\alpha)\,s ) 
  - \ln(- ( 1 - \alpha ) \,s )\big)^2 \nonumber\\
  &\quad-\big(\ln( (1-\alpha)\,s ) - \ln(- \alpha\,s )\big)\,
  \big(\ln(-(1-\alpha)\,s) - \ln( \alpha\,s )\big)\nonumber\\
  &\quad+\big(\ln(Q^2) -2\,\ln(-t_a)\big)\,\big(\ln( (1-\alpha)\,s ) 
  + \ln(- ( 1 - \alpha ) \,s ) - \ln(\alpha\,s) 
  - \ln(-\alpha\,s)\big) \nonumber\\
%%%%
  &-\frac{\bar H^a_T}{t\,t_b}
  \Big[
  \frac{1}{2}\big(\ln( (1-\alpha)\,s ) 
  - \ln(- ( 1 - \alpha ) \,s )\big)^2
  -\frac{1}{2}\big(\ln( \alpha\,s ) - \ln(- \alpha\,s )\big)^2
  \nonumber\\
  &\quad-\big(\ln( (1-\alpha)\,s ) - \ln(- \alpha\,s )\big)\,
  \big(\ln(-(1-\alpha)\,s) - \ln( \alpha\,s )\big)\nonumber\\
  &\quad+\big(\ln(Q^2) -2\,\ln(-t_b)\big)\,
  \big(\ln(\alpha\,s) + \ln(-\alpha\,s) - \ln( (1-\alpha)\,s ) 
  - \ln(- ( 1 - \alpha ) \,s )\big)\Big] \nonumber\\
%%%%
  &+2\frac{t_b\,H^a_T+t_a\,\bar H^a_T}{t (Q^2\,t + t_a\,t_b)}
  \Big[{\rm Li}_2\left(1+\frac{Q^2}{t_b}\right)  
  - {\rm Li}_2\left(1+\frac{Q^2}{t_a}\right)
  - {\rm Li}_2\left(1-\frac{t}{t_b}\right)
  + {\rm Li}_2\left(1-\frac{t}{t_a}\right)\nonumber\\
  &+\frac{1}{2}\ln^2(-t_b) - \frac{1}{2}\ln^2(-t_a) 
  -\ln(-t_b)\ln\frac{t}{t_b}
  +\ln(-t_a)\ln\frac{t}{t_a}
  \Big]  \nonumber\\
%%%%
  &-4\frac{H^a_T}{t\,t_a}\Big[\frac{\pi^2}{6} 
  - {\rm Li}_2\left(1+\frac{Q^2}{t_a}\right)\Big]
  +4\frac{\bar H^a_T}{t\,t_b}\Big[\frac{\pi^2}{6} 
  - {\rm Li}_2\left(1+\frac{Q^2}{t_b}\right)\Big]. 
\end{align}
We can easily observe that the logarithms in $s$ cancel out. Note,
however, the nontrivial phase structure: as we have discussed after
(\ref{NLOqqvertgg}), such a phase structure is expected, and in the
double-Regge limit ($t,Q^2 \ll M^2 \ll s)$ it will lead to the
anticipated decomposition. Finally we mention that, when adding the
two pentagon graphs, the remaining four lines in (\ref{pentpole}) and
the last two lines in (\ref{pentfin}) are simply multiplied by a
factor of $2$.

\section{Renormalization}

In this paper the Feynman gauge is adopted in all calculations. 
In order to regularize the singularities we have used the dimensional
regularization procedure. At this stage our results contain both
infrared and ultraviolet singularities: standard renormalization 
will remove the ultraviolet singularities, whereas the infrared
singularities will cancel only in the complete NLO result for the
photon impact factor \cite{fadinmartin}. In this section we will perform
the renormalization, and we will use the
modified minimal subtraction 
(${\overline{\rm MS}}$) scheme. In order to demonstrate the
cancellation of the ultraviolet $\epsilon$-poles, we first list the
ultraviolet divergencies of our diagrams in Figs.~3.1 and 3.4--11
(the box diagrams in Figs.~3.2 and 3.3, the pentagon graph Fig.~3.13, and
and the box diagrams Figs.~3.12 and 3.~14 are ultraviolet 
finite). Our analysis leads to the following results (deviating from
definitions in Eq. (\ref{eq:16}) we now include the coupling constant
$g$):
\begin{eqnarray}
A_{1}^{UV} = \frac{A^{(0)} g^4}{(4 \pi)^{2 - \epsilon}}
\frac{(-t)^{-\epsilon}}{\epsilon_{UV}} \frac{3 N_c}{2} \, , 
\end{eqnarray}
\begin{eqnarray}
A_{4}^{UV} = - \frac{A^{(0)} g^4}{(4 \pi)^{2 - \epsilon}}
\frac{(-t)^{-\epsilon}}{\epsilon_{UV}} \frac{1}{2 N_c} \, , 
\end{eqnarray}
\begin{eqnarray} 
A_5^{UV} = \frac{A^{(0)} g^4} 
{(4 \pi)^{2 - \epsilon}} \frac{(-t)^{-\epsilon}}{\epsilon_{UV}} 
\frac{3 N_c}{2}\; ,
\end{eqnarray}
\begin{eqnarray}
A_6^{UV} = - \frac{A^{(0)} g^4 }{(4 \pi)^{2 - \epsilon}}
\frac{(-t)^{-\epsilon}}{\epsilon_{UV}} \frac{1}{2 N_c}\,,
\end{eqnarray}
\begin{eqnarray}
A_{7+8+9}^{UV} = \frac{A^{(0)} g^4}{(4 \pi)^{2 - \epsilon}}
\frac{(-t)^{-\epsilon}}{\epsilon_{UV}} [\frac{5}{3}N_c - \frac{2}{3}
n_f ] \, , 
\end{eqnarray}
\begin{eqnarray}
A_{10}^{UV} = \frac{A^{(0)} g^4}{(4 \pi)^{2 - \epsilon}}
\frac{(-t_a)^{-\epsilon}}{\epsilon_{UV}} C_{\rm F} \, , 
\end{eqnarray}
\begin{eqnarray}
A_{11}^{UV} = - \frac{A^{(0)} g^4}{(4 \pi)^{2 - \epsilon}}
\frac{(-t_a)^{-\epsilon}}{\epsilon_{UV}} C_{\rm F} \, . 
\end{eqnarray}
The poles in $A_5$ and $A_6$, coming from the lower vertex
correction, as well as the gluon self-energy $A_7$ to $A_9$ can be
compared to standard textbook results.

In above formulas $g$, the strong coupling constant, denotes the 
unrenormalized one, the bare coupling. In order to perform the
usual renormalization, we make the replacement:
\begin{eqnarray}
g \rightarrow \frac{Z_1 g_r \mu^\epsilon}
{Z_2 \sqrt{Z_3}} \rightarrow g_r\Big[ 1 - \frac{\alpha_s}
{4\pi} \beta_0 \left( \frac{1} {\epsilon_{UV}} - \gamma_{\rm E} + \log(4 \pi)
\right) + \cdots \Big] \, ,
\end{eqnarray}
where $\beta_0 = (\frac{11}{6} N_c - \frac{1}{3} n_f)$, and 
$\gamma_{\rm E}$ is the Euler constant.
In the ${\overline{\rm MS}}$ scheme we have:
\begin{eqnarray}
Z_1 & = & 1 - \frac{\alpha_s}{4 \pi} (N_c + C_F) \Big[
\frac{1}{\epsilon_{UV}} - \gamma_{\rm E} + \log(4 \pi)\Big] \,, \\
Z_2 & = & 1 - \frac{\alpha_s}{4 \pi} C_F \Big[\frac{1}{\epsilon_{UV}} - 
\gamma_{\rm E} + \log(4 \pi)\Big]\,, \\
Z_3 & = & 1 + \frac{\alpha_s}{4 \pi} (\frac{5}{3} N_c - \frac{2}{3}
n_f) \Big[\frac{1}{\epsilon_{UV}} - \gamma_{\rm E} + \log(4 \pi) \Big] \,,  
\end{eqnarray}
for the vertex renormalization, the quark wave function
renormalization, and for the gluon wave function
renormalization, resp. Finally, we add the quark self-energy 
diagrams for the external legs (not shown in Fig.~3): 
\begin{eqnarray}
A_{\rm{quark\, self-energy}}^{UV} = - \frac{A^0\, g^4}{(4 \pi)^{2 -
\epsilon}} 
\frac{C_{\rm F}}{\epsilon_{\rm{UV}}}\,. 
\end{eqnarray}
It is easy to see that in this way all ultraviolet divergencies
cancel. In particular, the ultraviolet 
divergences in diagrams Figs.3.10 and 3.11 exactly cancel 
against each other as expected, similar to situation in the NLO calculation of 
$\gamma^* \rightarrow q \bar{q}$ in the $e^+e^-$ annihilation process.

\section{Conclusions}
In this paper we have calculated the high energy limit of the process
$\gamma^*q \to q\bar{q}\,q$ in next-to-leading order, and from the results we 
have extracted the NLO corrections to the coupling of the reggeized gluon 
to the vertex $\gamma^* \to q\bar{q}$. This calculation represents the first 
step in the computation of the NLO corrections to the photon impact factor. 
These NLO corrections will allow to perform a complete NLO analysis of the
BFKL prediction for the scattering process $\gamma^*\gamma^* \to \gamma^* \gamma^*$ at high energies.

In this paper we have listed the results of all one loop diagrams.
Using dimensional regularization we have carried out all loop
integrations, and our final results are expressed in terms of
logarithms and dilogarithms. We also show the explicit dependence upon
the helicities of the photon and the quarks. After renormalization our
results are free from ultraviolet divergencies, but they still contain
infrared singularities which will cancel once all NLO pieces of the
photon impact factor have been calculated and put together.

We have not yet attempted to combine the contributions from the Feynman 
diagrams into a single compact expression. The results for the individual
diagrams are sufficiently complicated and lengthy, and we found it useful to 
first list them separately. A closer investigation of the sum of all diagrams
will be presented in a forthcoming 
paper. This includes important consistency checks as well as investigations of
several special kinematic limits. 

\paragraph{Note:} shortly before this paper has been completed a short paper
(\texttt{hep-ph/0007119}) by V.\ Fadin, D.\ Ivanov, and M.\ Kotsky has
appeared which reports on a similar calculation of the $\gamma^*\to
q\bar{q}$ vertex.  However, the results presented in that paper are
written in terms of one-dimensional, two-dimensional, or even
three-dimensional integrals which have to be performed. We therefore
feel, at this stage, unable to make any comparison between our and
their results.

\pagebreak
\vskip 1.2cm
%%%%%%%%%%%%%%%%%%%%%%%%%%%%%%%%%%%%%%%%%%%%%%%%%%%%%%%%%%%%%%%%%%%%%%%%%%%
\centerline{\bf ACKNOWLEDGEMENTS}
\vskip 0.3cm      
We gratefully acknowledge a very helpful discussion with J.\ Collins. 

This work is partly Supported by the Graduiertenkolleg `Theoretische
Elementarteilchenphysik', by the Alexander von Humboldt foundation, by
the Graduiertenkolleg `Zuk\"unftige Entwicklungen der Teilchenphysik',
and by the TMR-Network `QCD and Particle Structure', contract number
FMRX-CT98-0194 (DG 12 - MIHT).
%%%%%%%%%%%%%%%%%%%%%%%%%%%%%%%%%%%%%%%%%%%%%%%%%%%%%%%%%%%%%%%%%%%%%%%%%%%
\section*{Appendix}
\begin{appendix}
\section{Basic functions}
In this section we define some basic functions, which are closely
related to scalar integrals and are built up from Logarithms and
Dilogarithms. The functions are defined in \cite{durham}. For
convenience we list them here.

The triangle functions depend on virtualities $p_i^2$ of the external
legs $i$, which we denote by $p_1^2, p_2^2$ and $p_3^2 = s_{12}$.  In
the case of the two mass triangle, where the second leg is on-shell,
$p_2^2=0$, we have only one simple logarithm:
\begin{equation}
  {\rm Lc}_0^{2m}(p^2_{1}, s_{12}) = \ln\frac{s_{12}}{p^2_{1}}. 
\end{equation}

For the three-mass triangle we have the following function 
\begin{equation}
  {\rm Lc}_0(p_1^2, p_2^2, s_{12}) =
  \frac{1}{\sqrt{-\Delta_3}}\left[
    \ln (a^+a^-) \ln\left(\frac{1-a^+}{1-a^-}\right)
    + 2{\rm Li}_2(a^+) - 2{\rm Li}_2(a^-)
  \right], 
\end{equation}
with the definitions 
\begin{align}
  \Delta_3 =& -p_{1}^4-p_{2}^4-s_{12}^2+2p_{1}^2 p_{2}^2 + 
   2 p_{1}^2 s_{12} + 2 p_{2}^2 s_{12}, \label{eq:gram3}\\
  a^\pm =& \frac{s_{12}+p_2^2-p_1^2\pm\sqrt{-\Delta_3}}{2 s_{12}}. 
\end{align}

The boxes depend on the virtualities of the external legs (we
put $p_4^2=s_{123}$) and on the invariants $s_{12} = (p_1+p_2)^2$,
$s_{23} = (p_2+p_3)^2$.  In the case of the one-mass box, with only the
fourth leg having a virtuality $s_{123} \neq 0$ we have the function
\begin{equation}
  {\rm Ld}^{1m}_0(s_{12}, s_{23}, s_{123}) = 
  {\rm Li}_2\left(1 - \frac{s_{12}}{s_{123}}\right)+ 
  {\rm Li}_2\left(1 - \frac{s_{23}}{s_{123}}\right)+ 
  \ln\frac{s_{12}}{s_{123}} \ln\frac{s_{23}}{s_{123}} - \frac{\pi^2}{6}. 
\end{equation}
For the box with the two adjacent legs "`1"' and "`4"' being off-shell,
($p_1^2\neq 0$ and $p_4^2=s_{123}\neq 0$) we have
\begin{equation}
  {\rm Ld}_0^a(s_{12}, s_{23}, p^2_{1}, s_{123}) = 
  {\rm Li}_2\left(1 - \frac{s_{12}}{s_{123}}\right)- 
  {\rm Li}_2\left(1 - \frac{p^2_{1}}{s_{12}}\right)+ 
  \frac{1}{2} \ln \frac{s_{12}^2}{p^2_{1} s_{123}}
  \ln\frac{s_{23}}{s_{123}}. 
\end{equation}
Finally, in case of the opposite box, where $p_2^2\neq0$ and
$p_4^2=s_{123}\neq 0$ we have
\begin{align}
  {\rm Ld}_0^{op}(s_{12}, s_{23}, p^2_{2}, s_{123}) =&\,  
  {\rm Li}_2\left(1-\frac{s_{12}}{s_{123}}\right)+ 
  {\rm Li}_2\left(1-\frac{s_{23}}{s_{123}}\right)
  -{\rm Li}_2\left(1-\frac{p^2_{2}}{s_{23}}\right)\nonumber\\
  &-{\rm Li}_2\left(1-\frac{p^2_{2}}{s_{12}}\right)+ 
  {\rm Li}_2\left(1-\frac{p^2_{2} s_{123}}{s_{12} s_{23}}\right)+
  \ln\frac{s_{12}}{s_{123}} \ln\frac{s_{23}}{s_{123}}.
\end{align}

The pentagon (in our case it always has one virtual external leg,
$p_5^2\neq 0$), will always be expressed in terms of these box
functions, since it is basically calculated in terms of boxes,
introduced by removing propagators from the pentagon.

\section{More Functions}
In the following we give a list of functions, appearing in our
calculations. These functions appear in the tensor decomposition of
the loop integrals and will recursively be expressed in terms of the
basic functions we introduced in the previous section.

\subsection{Functions for the two-mass triangle}
Here we list the functions which are neede in order to express the
tensor integrals of the two-mass triangle. The invariants are
explained above.
\begin{align}
  {\rm Lc}_{1}^{2m}(p_{1}^2, s_{12}) =& \,\frac{{\rm Lc}_{0}^{2m}(p_{1}^2, s_{12})}{s_{12} - p_{1}^2}\\ 
  {\rm Lc}_{2}^{2m}(p_{1}^2, s_{12}) =& -\frac{p_{1}^2 {\rm Lc}_{1}^{2m}(p_{1}^2, s_{12}) - 1}{s_{12} - p_{1}^2}\\ 
  {\rm Lc}_{3}^{2m}(p_{1}^2, s_{12}) =& -\frac{p_{1}^2 {\rm Lc}_{2}^{2m}(p_{1}^2, s_{12}) - 1/2}{s_{12} - p_{1}^2} 
\end{align}

\subsection{Three-mass triangle functions}
In the tensor decomposition of the three-mass triangle the following
functions appear (the Gram-determinant $\Delta_3$ is given in
eq.~(\ref{eq:gram3})): 

\begin{align}
  {\rm Lc}_{1}(p_{1}^2, p_{2}^2, s_{12}) =& \frac{1}{\Delta_3} \big[-2
  p_{1}^2 \ln\frac{s_{12}}{p_{1}^2} + (p_{1}^2 + p_{2}^2 - s_{12})
  \ln\frac{s_{12}}{p_{2}^2} \nonumber
 \\ &+ 
p_{1}^2 (s_{12} + p_{2}^2 - p_{1}^2) {\rm Lc}_{0}
(p_{1}^2, p_{2}^2, s_{12})\big]\\ 
{\rm Lc}_{2}(p_{1}^2, p_{2}^2, s_{12}) =& \frac{1}{2 \Delta_3} 
\big[ p_{1}^2 + p_{2}^2 - s_{12} - p_{2}^2 \ln\frac{s_{12}}{p_{2}^2} + 
  2 p_{2}^2 (p_{1}^2 - p_{2}^2 + s_{12}) {\rm Lc}_{1}(p_{1}^2,
  p_{2}^2, s_{12}) \nonumber\\&+ p_{1}^2 (-p_{1}^2 + p_{2}^2 + s_{12}) 
{\rm Lc}_{1}(p_{2}^2, p_{1}^2, s_{12}) - p_{1}^2 p_{2}^2 
{\rm Lc}_{0}(p_{1}^2, p_{2}^2, s_{12})\big]\\ 
  {\rm Lc}_{3}(p_{1}^2, p_{2}^2, s_{12}) =& \frac{1}{2 \Delta_3} 
\big[ -2 p_{1}^2 + (p_{2}^2 - s_{12}) \ln\frac{s_{12}}{p_{2}^2} + 
  p_{1}^4 {\rm Lc}_{0}(p_{1}^2, p_{2}^2, s_{12}) \nonumber\\&+ 3 p_{1}^2 (-p_{1}^2 + p_{2}^2 + s_{12}) {\rm Lc}_{1}(p_{1}^2, p_{2}^2, s_{12})\big]\\ 
  {\rm Lc}_{1S}(p_{1}^2, p_{2}^2, s_{12}) =& \frac{1}{2} \big[p_{1}^2 {\rm Lc}_{1}(p_{2}^2, p_{1}^2, s_{12}) + p_{2}^2 {\rm Lc}_{1}(p_{1}^2, p_{2}^2, s_{12})\big]\\ 
  {\rm Lc}_{2S}(p_{1}^2, p_{2}^2, s_{12}) =& \frac{1}{4 \Delta_3} \big[2 p_{1}^2 p_{2}^2 s_{12} {\rm Lc}_{1S}(p_{1}^2, p_{2}^2, s_{12})-
  \frac{1}{6} (p_{1}^4 (s_{12} + p_{2}^2 - p_{1}^2)
  \ln\frac{s_{12}}{p_{1}^2} \nonumber\\&+ 
  p_{2}^4 (s_{12} + p_{1}^2 - p_{2}^2) \ln\frac{s_{12}}{p_{2}^2} + 2 p_{1}^2 p_{2}^2 s_{12})\big]\\ 
  {\rm Lc}_{3S}(p_{1}^2, p_{2}^2, s_{12}) =& \frac{1}{6 \Delta_3} \big[2 p_{1}^2 p_{2}^2 s_{12} {\rm Lc}_{2S}(p_{1}^2, p_{2}^2, s_{12})- 
  \frac{1}{60} (p_{1}^6 (s_{12} + p_{2}^2 - p_{1}^2)
  \ln\frac{s_{12}}{p_{1}^2} \nonumber\\&+ 
  p_{2}^6 (s_{12} + p_{1}^2 - p_{2}^2) \ln\frac{s_{12}}{p_{2}^2} + p_{1}^2 p_{2}^2 s_{12} (p_{1}^2+p_{2}^2+s_{12})/2)\big]
\end{align}

\subsection{Adjacent Box}
For the box with two adjacent massive external legs we use the
abbreviation 
\begin{equation}
  \Delta_4 = 2 s_{23} \big[(s_{123}-s_{12})(s_{12}-p_{1}^2) - 
    s_{12} s_{23}\big].  
\end{equation}
With this we have the following functions related to
higher dimensionial scalar integrals:   
\begin{align}
  &{\rm Ld}_{1S}(s_{12}, s_{23}, p_{1}^2, s_{123}) = 
  -\frac{2 s_{12} s_{23}}{\Delta_4} 
  \big[{\rm Ld}_{0}^a(s_{12}, s_{23}, p_{1}^2, s_{123}) \nonumber\\&\qquad+ 
  \frac{1}{2} (s_{123} + p_{1}^2 - s_{23} - 
  2 \frac{p_{1}^2 s_{123}}{s_{12}}) {\rm Lc}_{0}(p_{1}^2, s_{23}, s_{123})
  \big]\\ 
  &{\rm Ld}_{2S}(s_{12}, s_{23}, p_{1}^2, s_{123}) = 
  - \frac{s_{12} s_{23}}{3 \Delta_4} \big[
  \frac{s_{23}}{2} \ln\frac{s_{123}}{s_{23}} + 
  s_{12} \ln\frac{s_{123}}{s_{12}} - 
  \frac{p_{1}^2}{2} \ln\frac{s_{123}}{p_{1}^2}
  \nonumber\\&\qquad 
  + s_{12} s_{23} {\rm Ld}_{1S}(s_{12}, s_{23}, p_{1}^2, s_{123}) +
  (s_{123} + p_{1}^2 - s_{23} - \frac{2 p_{1}^2 s_{123}}{s_{12}}) 
  {\rm Lc}_{1S}(p_{1}^2, s_{23}, s_{123}) \big]\\ 
  &{\rm Ld}_{3S}(s_{12}, s_{23}, p_{1}^2, s_{123}) = 
  -\frac{s_{12} s_{23}}{5 \Delta_4} 
  \big[
  \frac{s_{23}^2}{24} \ln\frac{s_{123}}{s_{23}} 
  + \frac{s_{12}^2}{12}\ln\frac{s_{123}}{s_{12}} 
  - \frac{p_{1}^4}{24} \ln\frac{s_{123}}{p_{1}^2} 
  + \frac{s_{12} s_{23}}{12}
  \nonumber\\&\qquad+
  s_{12} s_{23} {\rm Ld}_{2S}(s_{12}, s_{23}, p_{1}^2, s_{123}) + 
  (s_{123} + p_{1}^2 - s_{23} - \frac{2 p_{1}^2
   s_{123}}{s_{12}}) {\rm Lc}_{2S}(p_{1}^2, s_{23}, s_{123})
  \big]
\end{align}
In the tensor decomposition of adjacent box integrals the
following functions are used:  
\begin{align}
  &{\rm Ld}_{1}(s_{12}, s_{23}, p_{1}^2, s_{123}) = 
    -\big[{\rm Ld}_{1S}(s_{12}, s_{23}, p_{1}^2, s_{123}) 
    + {\rm Lc}_{0}(p_{1}^2, s_{23}, s_{123})\big]\\
  &{\rm Ld}_{21}(s_{12}, s_{23}, p_{1}^2, s_{123}) = 
    -\frac{2}{s_{12}} \big[3 {\rm Ld}_{2S}(s_{12}, s_{23}, p_{1}^2, s_{123}) + 
    {\rm Lc}_{1S}(p_{1}^2, s_{23}, s_{123})\big] \nonumber\\&\qquad 
    - {\rm Lc}_{1}(s_{23}, p_{1}^2, s_{123})\\ 
  &{\rm Ld}_{22}(s_{12}, s_{23}, p_{1}^2, s_{123}) = 
    2 \frac{s_{123} - s_{12}}{s_{12} s_{23}} 
    \big[3 {\rm Ld}_{2S}(s_{12}, s_{23}, p_{1}^2, s_{123}) 
    + {\rm Lc}_{1S}(p_{1}^2, s_{23}, s_{123})\big] \nonumber\\&\qquad- 
    \frac{s_{12}}{s_{23}} {\rm Lc}_{1}^{2m}(p_{1}^2, s_{12}) + 
      \frac{s_{123}}{s_{23}} {\rm Lc}_{1}(s_{23}, p_{1}^2, s_{123})\\ 
  &{\rm Ld}_{24}(s_{12}, s_{23}, p_{1}^2, s_{123}) = 
    2 \frac{p_{1}^2 - s_{12}}{s_{12} s_{23}} 
    \big[3 {\rm Ld}_{2S}(s_{12}, s_{23}, p_{1}^2, s_{123}) 
    + {\rm Lc}_{1S}(p_{1}^2, s_{23}, s_{123})\big] \nonumber\\&\qquad- 
    \frac{s_{12}}{s_{23}} {\rm Lc}_{1}^{2m}(s_{12}, s_{123}) + 
    \frac{p_{1}^2}{s_{23}} {\rm Lc}_{1}(s_{23}, p_{1}^2, s_{123})\\
  &{\rm Ld}_{311}(s_{12}, s_{23}, p_{1}^2, s_{123}) = 
  -\Big(\frac{12}{s_{12}^2} 
  \big[5 {\rm Ld}_{3S}(s_{12}, s_{23}, p_{1}^2, s_{123}) + 
  {\rm Lc}_{2S}(p_{1}^2, s_{23}, s_{123})\big] \nonumber\\&\qquad+ 
  \frac{s_{12} + p_{1}^2}{s_{12}} {\rm Lc}_{3}(s_{23}, p_{1}^2, s_{123}) 
  + \frac{s_{23}}{s_{12}} {\rm Lc}_{2}(p_{1}^2, s_{23}, s_{123})\Big)\\ 
&{\rm Ld}_{314}(s_{12}, s_{23}, p_{1}^2, s_{123}) = 
  12 \frac{p_{1}^2 - s_{12}}{s_{12}^2 s_{23}} 
  \big[5 {\rm Ld}_{3S}(s_{12}, s_{23}, p_{1}^2, s_{123}) + 
  {\rm Lc}_{2S}(p_{1}^2, s_{23}, s_{123})\big] \nonumber\\&\qquad+ 
  \frac{p_{1}^4}{s_{12} s_{23}} {\rm Lc}_{3}(s_{23}, p_{1}^2, s_{123}) - 
  \frac{s_{12} + p_{1}^2}{s_{12}} {\rm Lc}_{2}(p_{1}^2, s_{23}, s_{123}) - 
  \frac{s_{12}}{2 s_{23}} {\rm Lc}_{1}^{2m}(s_{12}, s_{123})\\ 
&{\rm Ld}_{322}(s_{12}, s_{23}, p_{1}^2, s_{123}) = 
  -\Big( 12 \frac{(s_{123} - s_{12})^2}{s_{12}^2 s_{23}^2} 
  \big[5 {\rm Ld}_{3S}(s_{12}, s_{23}, p_{1}^2, s_{123}) 
  + {\rm Lc}_{2S}(p_{1}^2, s_{23}, s_{123})\big] \nonumber\\&\qquad+ 
  s_{123} \frac{s_{12} s_{123} + p_{1}^2 s_{123} - 
    2 p_{1}^2 s_{12}}{s_{12} s_{23}^2} {\rm Lc}_{3}(s_{23}, p_{1}^2, s_{123}) - 
  \frac{s_{12}}{2 s_{23}} {\rm Lc}_{2}^{2m}(p_{1}^2, s_{12})
\nonumber\\&\qquad+ 
    s_{12} \frac{s_{12} - s_{123}}{2 s_{23}^2} {\rm Lc}_{1}^{2m}(s_{12},
s_{123}) + 
    s_{123} \frac{s_{123} - 2 s_{12}}{s_{12} s_{23}} {\rm Lc}_{2}(p_{1}^2, s_{23}, s_{123}) \nonumber\\&\qquad+ 
    \frac{p_{1}^2 s_{12} + s_{12} s_{23} - s_{12} s_{123}}{2 s_{23}^2}
{\rm Lc}_{1}^{2m}(p_{1}^2, s_{12})\Big)%\\
\end{align}
\begin{align}
&{\rm Ld}_{344}(s_{12}, s_{23}, p_{1}^2, s_{123}) = 
  -\Big( 12 \frac{(s_{12} - p_{1}^2)^2}{s_{12}^2 s_{23}^2} 
  \big[5 {\rm Ld}_{3S}(s_{12}, s_{23}, p_{1}^2, s_{123}) + 
  {\rm Lc}_{2S}(p_{1}^2, s_{23}, s_{123})\big] \nonumber\\&\qquad+ 
  p_{1}^2 \frac{p_{1}^2 - 2 s_{12}}{s_{12} s_{23}} {\rm Lc}_{2}(p_{1}^2, s_{23}, s_{123}) +
  p_{1}^4 \frac{p_{1}^2 - s_{12}}{s_{12} s_{23}^2} {\rm Lc}_{3}(s_{23}, p_{1}^2, s_{123}) \nonumber\\&\qquad+ 
  s_{12} \frac{s_{12} - p_{1}^2}{2 s_{23}^2} {\rm Lc}_{1}^{2m}(s_{12},
s_{123}) + 
  \frac{s_{12}}{2 s_{23}} {\rm Lc}_{2}^{2m}(s_{12}, s_{123})\Big)
\end{align}
\subsection{One-mass box}
For box integrals with one massive external leg we use the
abbreviation 
\begin{equation}
  \Delta_4^{1m} = 2 s_{12} s_{23} (s_{123} - s_{12} - s_{23}). 
\end{equation}
The functions we are using for tensor integrals are: 
\begin{align}
  {\rm Ld}_{1S}^{1m}(s_{12}, s_{23}, s_{123}) &=
     -\frac{2 s_{12} s_{23}}{\Delta_4^{1m}} 
     {\rm Ld}_0^{1m}(s_{12}, s_{23}, s_{123})\\ 
  {\rm Ld}^{1m}_{2S}(s_{12}, s_{23}, s_{123}) &=
    - \frac{s_{12} s_{23}}{3 \Delta_4^{1m}} \big[s_{12} s_{23} 
    {\rm Ld}^{1m}_{1S}(s_{12}, s_{23}, s_{123}) 
    + s_{23} \ln\frac{s_{123}}{s_{23}} + s_{12} \ln\frac{s_{123}}{s_{12}}\big]\\ 
  {\rm Ld}^{1m}_{1}(s_{12}, s_{23}, s_{123}) &=
    -{\rm Ld}^{1m}_{1S}(s_{12}, s_{23}, s_{123})\\
  {\rm Ld}^{1m}_{21}(s_{12}, s_{23}, s_{123}) &=- \frac{1}{s_{12}}
    \big[6 {\rm Ld}^{1m}_{2S}(s_{12}, s_{23}, s_{123}) + 
    s_{23} {\rm Lc}_{1}^{2m}(s_{23}, s_{123})\big]\\ 
  {\rm Ld}^{1m}_{22}(s_{12}, s_{23}, s_{123}) &=
    {\rm Ld}^{1m}_{1}(s_{12}, s_{23}, s_{123}) 
    - {\rm Ld}^{1m}_{21}(s_{12}, s_{23}, s_{123}).
\end{align}
\end{appendix}

%%% Local Variables: 
%%% mode: latex
%%% TeX-master: t
%%% End: 

%%%%%%%%%%%%%%%%%%%%

%%%%%%%%%%%%%%%%%%%%%%%%%%%%%%%%%%%%%%%%%%%%%%%%%%%%%%%%%%%%%%%%%%%%%%%%%%%

\newpage 
\begin{thebibliography}{99}
\bibitem{bfkl} E.A.\ Kuraev, L.N.\ Lipatov, V.S.\ Fadin, Sov.\ Phys.\ JETP
  {\bf{45}} (1977) 199; Ya.Ya.\ Balitskii and L.N.\ Lipatov, Sov.\ J.\
  Nucl.\  Phys.\ {\bf 28} (1978) 822.
  
\bibitem{bartels2} J.\ Bartels, A.\ De Roeck, H.\ Lotter, Phys.\ 
  Lett.\ {\bf B 389} (1996) 742.
  
\bibitem{brodsky} S.J.\ Brodsky, F.\ Hautmann, and D.E.\ Soper, Phys.\ 
  Rev.\ {\bf D56} (1997) 6957; Phys.\ Rev.\ Lett.\ {\bf 78} (1997)
  803.
  
\bibitem{bartels1} J.\ Bartels, C.\ Ewerz, and R.\ Staritzbichler,
  \texttt{hep-ph/0004029}.
  
\bibitem{DSR} A.\ Donnachie, S.\ S{\"o}ldner-Rembold,
  \texttt{hep-ph/0001035}, to be published in the proceedings of the
  UK Phenomenology Workshop on Collider Physics, Durham 1999.
  
\bibitem{NLOint} S.J.\ Brodsky, V.S.\ Fadin, V.T.\ Kim, L.N.\ Lipatov,
  and G.B.\ Pivovarov, JETP Lett.\ {\bf 70} (1999) 155.

\bibitem{FL} V.S.\ Fadin, L.N.\ Lipatov, Phys.\ Lett.\ {\bf B 429} (1998)
 127 and references therein.

\bibitem{CC} M.\ Ciafaloni, G.\ Camici, Phys.\ Lett.\ {\bf B 430} (1998)
 349 and references therein.
 
\bibitem{wavefun} A.H.\ Mueller, Nucl.\ Phys.\ {\bf B335} (1990) 115;
  N.\ Nikolaev, B.G.\ Zakharov, Z.\ Phys.\ {\bf C49} (1991) 607, {\bf
    C53} (1992) 331; S.J.\ Brodsky, P.\ Hoyer, and L.\ Magnea, Phys.\ 
  Rev.\ {\bf D55} (1997) 5585; F.\ Hautmann, Z.\ Kunszt, D.E.\ Soper,
  Phys.\ Rev.\ Lett.\ {\bf 81} (1998) 3333; D.\ Yu.\ Ivanov, M.\ 
  W{\"u}sthoff, Eur.\ Phys.\ J.\ {\bf C8} (1999) 107; S.\ Gieseke and
  C.F.\ Qiao, Phys.\ Rev.\ {\bf D61} (2000) 074028.
  
\bibitem{fadin2} V.S.\ Fadin, R.\ Fiore, and A.\ Quartarolo, Phys.\ 
  Rev.\ {\bf D50} (1994) 2265.
  
\bibitem{bartels3} J.\ Bartels, Nucl.\ Phys.\ {\bf B 175} (1980) 365.
  
\bibitem{BGQ2} J.\ Bartels, S.\ Gieseke, C.-F.\ Qiao, in preparation

\bibitem{bern} Z.\ Bern, L.\ Dixon, and D.A.\ Kosower, Nucl.\ Phys.\ 
  {\bf B 412} (1994) 751 and references therein.
 
\bibitem{durham} J.M.\ Campbell, E.W.N.\ Glover, and D.J.\ Miller,
  Nucl.\ Phys.\ {\bf B 498} (1997) 397.

\bibitem{feyncalc}
R.~Mertig, M.~B{\"o}hm and A.~Denner, 
Comput.\ Phys.\ Commun.\  {\bf 64}, 345 (1991).

\bibitem{fadinmartin} V.S.\ Fadin, A.D.\ Martin, Phys.\ Rev.\ {\bf
    D60} (1999) 114008.
\end{thebibliography}
\end{document}